\documentclass[referee]{gji}
\usepackage{flushend}
\usepackage{timet}
\usepackage{color}
\usepackage{balance}
\usepackage{amsmath}    
\usepackage{amssymb}    
\usepackage{mathrsfs}  
\usepackage{graphicx}
\usepackage{subfigure}
\usepackage{epstopdf}
\usepackage{ulem}

\newcommand{\ts}[1]{\boldsymbol {\mathsf #1}}
\newcommand{\dd}[1]{\, {\rm d} #1}
\newcommand{\co}[1]{\color{blue} #1}

\usepackage{amssymb}
\makeatletter
\newcommand{\rmnum}[1]{\romannumeral #1}
\newcommand{\Rmnum}[1]{\expandafter\@slowromancap\romannumeral #1@}
\allowdisplaybreaks[4]
\title{Exact closed-form solutions for Lamb's problem (III):\\
 The case for buried source and receiver}
\author[Xi Feng and Haiming Zhang]{Xi Feng and Haiming Zhang \\
  Department of Geophysics, School of Earth and Space Sciences, Peking University \emph{100871}, P. R. China.\\
  Email: zhanghm@pku.edu.cn}
\date{In original form 2020}
\pagerange{\pageref{firstpage}--\pageref{lastpage}}
\volume{200}
\pubyear{2020}

\graphicspath{{fig/}}

\begin{document}

\label{firstpage}

\maketitle

\begin{summary}
In this article, we derive the exact closed-form solution for the displacement in the interior of an
elastic half-space due to a buried point force 
with Heaviside step function time history.
It is referred to as the tensor Green's function for the elastic wave equation in a uniform half-space, 
also a natural generalization of the classical 3-D Lamb's problem, 
for which previous solutions have been restricted to the cases of either the source or the receiver or both are located on the free surface. 
Starting from the complex integral solutions of Johnson, 
we follow the similar procedures presented by Feng and Zhang to obtain the closed-form expressions in terms of elementary functions as well as elliptic integrals. 
Numerical results obtained from our closed-form expressions agree perfectly with those of Johnson, 
which validates our explicit formulae conclusively.
\end{summary}
\section{Introduction}
Lamb's problem, 
which describes the transient response on the free surface of an elastic half-space due to a time-dependent point force, 
is a major topic which has extensively been treated in the classical literature in theoretical seismology. 
Solution to Lamb's problem has become a cornerstone of many elastic wave propagation problems, 
and has been widely used in both scientific and engineering fields. 
The subject was first introduced by Lamb ({\co 1904}), 
who derived the expression for the surface displacement excited by 
a vertical point or line force with nearly impulsive time dependence. 
The monochromatic solutions due to a time-harmonic force was first obtained and superposed to transient field via Fourier synthesis.  
Following Lamb's method, Nakano ({\co 1925}) considered the surface displacement caused by a buried line force, 
and found for the first time a diffracted wave arriving between the P and S waves after solving the complex integrals asymptotically by the steepest descent method under the stationary phase assumption.
Lapwood ({\co 1949}) discussed eight types of solutions near the surface by the line from a line force by deforming the integration contours on the Riemann surface.

On the other hand, Cagniard ({\co 1939}) developed an alternative approach based on the Laplace transform with respect to the temporal variable. 
Through clever algebraic manipulation, he succeeded in writing the integral in the wavenumber domain into the form of the Laplace transform integral, 
from which the solution in the time domain can be recognized by a mere inspection instead of going through the inverse Laplace transform. Following Cagniard's method, Garvin ({\co 1956}) obtained the closed-form expression for the surface displacement elicited by a line of impulsive pressure when the source is buried and the displacements are sought at
the surface. This problem was revisited by S\'anchez-Sesma \textit{et. al.} ({\co 2013}) who presented a generalization
to Garvin’s problem when both the source and the receiver are placed at an arbitrary location, that is,
for a type III problem, albeit in two dimensions. Also, Pinney (1954) considered the response
due to impulsive S (torsional) and P (voluminal) point sources when $\lambda=\mu$. 
However, Cagniard's (1939) abstruse mathematical treatment
eluded most researchers of the time until de Hoop's (1960) contribution greatly simplified 
the algebra and made the derivation much more accessible to seismologists.
The method is now referred to as the \textit{Cagniard-de Hoop method}, which became the most effective tool to handle Lamb's problem. 

For the case of a point force in 3-D space, we classify Lamb's problem into three kinds according to the locations of the source and the receiver. Lamb's problem of the first kind refers to the case when both the source and the receiver are situated on the free surface. 
Pekeris ({\co 1955a}) considered the problem of a vertical force varying with a time history of the Heaviside step function. 
A closed-form solution was obtained by an approach equivalent to that of Cagniard ({\co 1939}). 
The displacement from a tangential point force was derived by Chao ({\co 1960}). 
In both Pekeris ({\co 1955a}) and Chao ({\co 1960}) the Poisson's ratio of the elastic medium was 0.25, 
which is an adequate assumption for the Earth. 
The restriction was removed by Mooney ({\co 1974}), 
who extended Pekeris's integral form solution to arbitrary Poisson's ratio. 
Richards ({\co 1979}) gave a complete set of exact formulae for both vertical and tangential forces and for arbitrary Poisson's ratio. 
Kausel ({\co 2012}) revisited the problem and his expressions are much simpler than those of Richards ({\co 1979}), only containing elementary functions and elliptic integrals. 
At present, Lamb's problem of the first kind is considered to have been fully solved.

When either the source or the receiver is buried beneath the surface, the problem becomes Lamb's problem of the second kind. 
For a buried vertical force, the exact solution for the motion on the surface was given in Pekeris ({\co 1955b}), 
followed by the corresponding numerical results in Pekeris and Lifson ({\co 1957}), 
where basic properties of the wavefield were discussed. 
A similar approach was adopted by Aggarwal and Ablow ({\co 1967}) to obtain the exact solution expressed in terms of finite integrals appropriate for numerical evaluations, 
and an asymptotic formula for the Rayleigh wave motion was also derived. Eason ({\co 1966}) investigated the interior response in the half-space when the source is placed on the free surface. 
By virtual of the reciprocity theorem, the problem is equivalent to the
case of solving for the motion on the free surface due to a buried force. The direct Laplace inversion was used instead of the Cagniard-de Hoop method in Eason ({\co 1966}), and the final results were quite complicated. 
Analytical expressions for the surface displacements due to a buried dislocation source were presented by Kawasaki ({\co 1972a, b}) together with numerical results, 
and a more straightforward derivation was obtained by Sato ({\co 1973}).

Johnson ({\co 1974}), hereafter referred to as J74, derived the complete solution for the surface displacements generated by a point force in the half-space using the Cagniard-de Hoop method.
Nine components of the Green's function for three orthogonal directions of displacements due to three orthogonal directions of the forces were expressed in terms of definite integrals convenient for numerical calculations.
The formulae for spatial derivatives of the Green's function were also presented, for the application of the representation theorem. 
Emani and Eskandari-Ghadi ({\co 2019}) obtained a full set of solutions
in a simple and elegant form using a slightly modified version of the Cagniard-de Hoop method and presented the derivation of the final solutions in great detail.
Feng and Zhang ({\co 2018}), hereafter referred to as FZ18, reported a mathematical procedure to cast the formulae in J74 into closed-form expressions, which contain both elementary functions and three kinds of elliptic integrals. 
However, the formulae are valid when the Poisson's ratio is less than 0.2631. 

Because of the mathematical complexity, few researches 
have considered the Lamb's problem when the source and the receiver are both buried beneath the surface. As a natural extension of the traditional Lamb's problem, this is known as Lamb's problem of the third kind, which is the focus of the current study. 
J74 has provided the complete integral solutions, 
but in complex integral form difficult for numerical calculations. 
Here, we first rewrite the expressions of J74 
into a simpler form suitable for numerical calculation 
by a change of variable in the integration.
Then, we follow the procedures similar to those in FZ18 to obtain the closed-form solutions expressed in terms of elementary algebraic functions as well as elliptic integrals. Our results agree perfectly with the numerical results of J74, which validates conclusively our explicit formulae. 
Our results are obviously advantageous over other solutions expressed in integral forms 
when specific types of seismic waves are to be analyzed.
The solution obtained in this study may also
play a key role in the analysis of dynamic rupture on a complex shaped fault in 3-D half-space (Zhang and Chen, {\co 2006}). 

We start in Section \ref{Section configuration} by introducing the geometry of the problem and definitions of notations used in derivation. 
In Section \ref{Section johnson} integral form results obtained by J74 are introduced. The derivation method to obtain the equivalent closed-form solution is explained in Section \ref{Section derivation method}. 
In Section \ref{Section reflected wave1} and \ref{Section reflected wave2}, different types of waves are analyzed separately. The 
ultimate results are summarized in Section \ref{Section guide}.
Our formulae are validated by comparison of numerical results with those of J74 in Section \ref{Section Numerical Results}
with further discussion on the Rayleigh wave part of the Green's function.
\section{Geometry of the problem and definitions of notations}\label{Section configuration}
\begin{figure}
\centering
\includegraphics[width=.6\textwidth]
{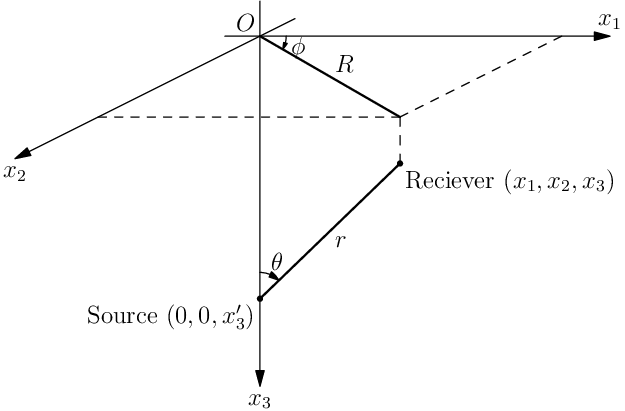}
\caption{
Geometry for the solution of Lamb's problem of the third kind.
The coordinates of the source and the receiver are ($x_{1}$, $x_{2}$, $x_{3}$) and ($0$, $0$, $x'_{3}$), respectively. $r$ denotes the distance between the source and the receiver, and $R$ denotes the epicentral distance, i.e., the projection of $r$ in the horizontal plane. $\theta$ is the angle
between the negative $x_3$ axis and the ray from the source to the receiver, and $\phi$ is the angle between $R$ and $x_1$ axis.}
\label{Figure geometry}
\end{figure}

Our aim is to derive the solution of Lamb's problem of the third kind, i.e., the displacement field excited by a force 
with a Heaviside step function time dependence
acting at some point beneath the surface of the elastic homogeneous half-space with the Lam\'{e} constants $\lambda$ and $\mu$ and the density $\rho$. A Cartesian coordinate system is established such that the free surface coincides with the plane $x_3=0$, and the positive $x_3$ axis points downward (Figure \ref{Figure geometry}).
The source is located on the $x_3$ axis with a depth $x_3'$, and the receiver located at an arbitrary point with coordinates $(x_1, x_2, x_3)$. The vertical displacement is positive down.
The following is a list of notations used throughout the paper:\\
\begin{tabular}{|c|l|}
\hline
$\lambda$, \,$\mu$ & Lam\'{e} constants \\
$\rho$ & mass density \\
$\alpha$ & velocity of P wave \\
$\beta$ & velocity of S wave \\
$k=\frac{\alpha}{\beta}$ & ratio of P to S wave velocity \\
$\kappa$ & $\sqrt{k^2-1}$ \\
$t$ & time \\
$T_{\rm dP}=\frac{\alpha t}{r}$ & dimensionless time in $\ts{G}^{\rm dP}$\\
$T_{\rm dS}=\frac{\beta t}{r}$ & dimensionless time in $\ts{G}^{\rm dS}$\\
$T_{\rm PS}=\frac{\alpha t}{R}$ & dimensionless time in $\ts{G}^{\rm PS}$\\
$t_{\rm PS}$ & arrival time of PS wave\\
\hline
\end{tabular}
\begin{tabular}{|c|l|}
\hline
$x'_{3}$ & depth of source\\
($x_{1}$, $x_{2}$, $x_{3}$) & coordinates of receiver\\
$z$ & $x_{3}/R$\\
$z'$ & $x'_{3}/R$\\
$r$ & distance between source and receiver \\
$R$ & epicentral distance between source and receiver \\
$\theta$ & angle
between the negative $x_3$ axis and $r$\\
$\phi$ & azimuth\\
$\mathcal{H}(\cdot)$ & Heaviside step function \\
${\rm Re}(\cdot)$ & real part \\
${\rm Im}(\cdot)$ & imaginary part \\
\hline
\end{tabular}
\section{Main results from J74}\label{Section johnson}
\begin{figure}
\centering
\subfigure{
\begin{minipage}{.48\textwidth}
\centering
\includegraphics[scale=0.75]
{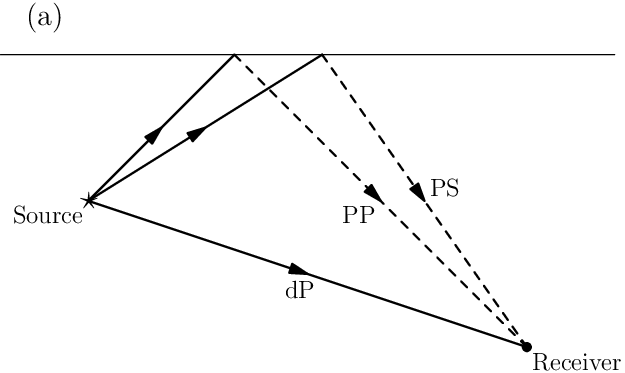}
\end{minipage}
}
\subfigure{
\begin{minipage}{.48\textwidth}
\centering
\includegraphics[scale=0.75]
{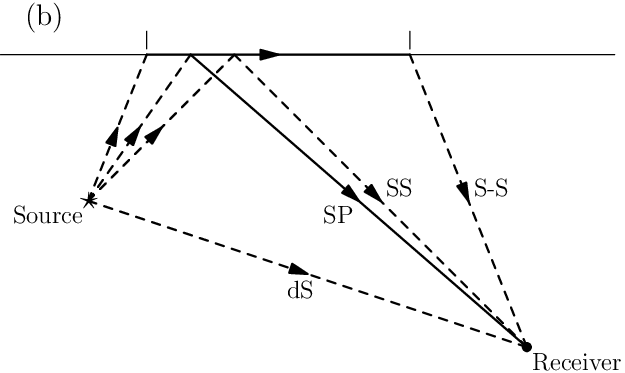}
\end{minipage}
}
\caption{
Schematic ray paths for 
(a) direct P wave (dP), reflected PP and PS waves; 
(b) direct S wave (dS), reflected SS and SP waves, and the surface refracted S-S wave with a segment of evanescent P wave on the free surface. 
P and S paths are marked in solid and dashed lines, respectively.
}
\label{Figure wave phase}
\end{figure}

The solutions for Lamb's problem of the third kind were given by
J74 in terms of integral forms for the Green's function, which consists of seven parts:
\begin{align}\label{Gfull}
\ts{G}=
\ts{G}^{\rm dP}+\ts{G}^{\rm dS}+
\ts{G}^{\rm PP}+\ts{G}^{\rm SS}
+\ts{G}^{\rm PS}+\ts{G}^{\rm SP}
+\ts{G}^{\rm S\text{-}S},
\end{align}
the tensor $\ts{G}$ with superscripts $\rm dP$, $\rm dS$, $\rm PP$, $\rm SS$, $\rm PS$, $\rm SP$ and $\rm S\text{-}S$ stand for tensor Green's functions for the direct P and S waves, the surface reflected PP, SS, SP and PS waves, and the surface refracted S-S wave, respectively, of which the schematic ray paths are shown in Figure \ref{Figure wave phase}. 
The analytical expressions for the direct P and S waves were given in eqs. (39)--(43) in J74. 
For completeness we quote his results here using our notations:
\begin{align}
\label{GdP}
\ts{G}^{\rm dP}&=\frac{\mathcal{H}\left(T_{\rm dP}-1\right)}{8\pi\rho\alpha^2 r}
\ts{M}^{\rm dP},
\\
\label{GdS}
\ts{G}^{\rm dS}&=\frac{\mathcal{H}\left(T_{\rm dS}-1\right)}{8\pi\rho\beta^2 r}
\ts{N}^{\rm dS},
\end{align}
where
\begin{align}\label{GdP component}
\notag
M_{11}^{\rm dP}=&(3T_{\rm dP}^2-1)\sin^2\theta\cos^2\phi-(T_{\rm dP}^2-1),
\\
\notag
M_{21}^{\rm dP}=M_{12}^{\rm dP}=&(3T_{\rm dP}^2-1)\sin^2\theta\sin\phi\cos\phi,
\\
M_{22}^{\rm dP}=&(3T_{\rm dP}^2-1)\sin^2\theta\sin^2\phi-(T_{\rm dP}^2-1),
\\
\notag
M_{31}^{\rm dP}=M_{13}^{\rm dP}=&-(3T_{\rm dP}^2-1)\sin\theta\cos\theta\cos\phi,
\\
\notag
M_{32}^{\rm dP}=M_{23}^{\rm dP}=&-(3T_{\rm dP}^2-1)\sin\theta\cos\theta\sin\phi,
\\
\notag
M_{33}^{\rm dP}=&(3T_{\rm dP}^2-1)\cos^2\theta-(T_{\rm dP}^2-1),
\end{align}
and 
\begin{align}\label{GdS component}
\notag
N_{11}^{\rm dS}=&-(3T_{\rm dS}^2-1)\sin^2\theta\cos^2\phi+(T_{\rm dS}^2+1),
\\
\notag
N_{21}^{\rm dS}=N_{12}^{\rm dS}=&-(3T_{\rm dS}^2-1)\sin^2\theta\sin\phi\cos\phi,
\\
N_{22}^{\rm dS}=&-(3T_{\rm dS}^2-1)\sin^2\theta\sin^2\phi+(T_{\rm dS}^2+1),
\\
\notag
N_{31}^{\rm dS}=N_{13}^{\rm dS}=&(3T_{\rm dS}^2-1)\sin\theta\cos\theta\cos\phi,
\\
\notag
N_{32}^{\rm dS}=N_{23}^{\rm dS}=&(3T_{\rm dS}^2-1)\sin\theta\cos\theta\sin\phi,
\\
\notag
N_{33}^{\rm dS}=&(3T_{\rm dS}^2-1)\sin^2\theta-2(T_{\rm dS}^2-1).
\end{align}
The direct P and S waves taken together represent the Green's function for a homogeneous elastic whole space. 

The expression for $\ts{G}^{\rm PS}$ is given in a complex integral form in J74 as follows:
\begin{equation}\label{PS Johnson}
\ts{G}^{\rm PS}=\frac{1}{2\pi^2\rho}\int^{p_u}_0\mathcal{H}\left(t-t_{\rm PS}\right){\rm Re}\left[{\rm i}\sigma^{-1}\left(R+q\left(x'_{3}/\eta_{\alpha}+x_3/\eta_{\beta}\right)\right)^{-1}\ts{N}^{\rm PS}\right]{\rm d}p,
\end{equation}
where
\begin{align}
\ts{N}^{\rm PS}=\begin{bmatrix}
4\gamma\eta_{\beta}\left(\left(q^2+p^2\right)\cos^2{\phi}-p^2\right) &
4\gamma\eta_{\beta}\left(q^2+p^2\right)\sin{\phi}\cos{\phi} &
4q\gamma\eta_{\alpha}\eta_{\beta}\cos{\phi}\\
4\gamma\eta_{\beta}\left(q^2+p^2\right)\sin{\phi}\cos{\phi} &
4\gamma\eta_{\beta}\left(\left(q^2+p^2\right)\sin^2{\phi}-p^2\right) &
4q\gamma\eta_{\alpha}\eta_{\beta}\sin{\phi}\\
4q\gamma\left(q^2-p^2\right)\cos{\phi} &
4q\gamma\left(q^2-p^2\right)\sin{\phi} &
4\gamma\eta_{\alpha}\left(q^2-p^2\right) 
\end{bmatrix},
\end{align}
and
\begin{align}
\notag
&
\eta_\alpha=\sqrt{\alpha^{-2}+p^2-q^2}, \ {\rm Re}(\eta_\alpha)\geq 0
&&
\eta_{_\beta}=\sqrt{\beta^{-2}+p^2-q^2}, \ {\rm Re}(\eta_\beta)\geq 0
\\
&
\gamma=\eta_{_\beta}^2+p^2-q^2,
&&
q=(x'_3\eta_{\alpha}+x\eta_{\beta}-t)/R,
\\
\notag
&
\sigma=\left(\eta_{_\beta}^2+p^2-q^2\right)^2+4\eta_\alpha\eta_{_\beta}\left(q^2-p^2\right).
\end{align}
J74 presented a complicated procedure to calculate $p_u$ and $t_{\rm PS}$ because no explicit expressions for them exist. We will present two simple methods to calculate $p_u$ in Section \ref{Section reflected wave1} and $t_{\rm PS}$ in Appendix \ref{Section tps}, respectively.
The derivation of closed-form formulae for $\ts{G}^{\rm PS}$ and $\ts{G}^{\rm SP}$ will be presented in Section \ref{Section reflected wave1}. 
The expressions for $\ts{G}^{\rm SP}$ can be directly obtained from those for $\ts{G}^{\rm PS}$ by exchanging the positions of $x'_3$ and $x_3$.
The expressions for the other reflected waves $\ts{G}^{\rm PP}$ and $\ts{G}^{\rm SS}$, together with the surface refracted S-S wave $\ts{G}^{\rm S\text{-}S}$, are quite similar to the solutions for Lamb's problem of the second kind, which were also presented in J74. 
Therefore, the corresponding closed-form results can be obtained by a straightforward application of the technique in FZ18, 
which will be discussed in Section \ref{Section reflected wave2}. 
\section{Derivation Method}\label{Section derivation method}
In this section, the general method to 
convert the formulae in J74 into closed-form expressions is explained from a mathematical point of view.
It is based on the following theorem (Armitage and Eberlein, {\co 2006}):

\textit{An integral of the type $\displaystyle \int R(x)\frac{{\mathrm d}x}{\sqrt{W(x)}}$, where $W(x)$ is a quartic polynomial in $x$ with no real roots, and $R$ denotes a rational fraction of $x$, is called an elliptic integral, which can be expressed as a finite sum of elementary functions and the three kinds of standard elliptic integrals.}

The calculation of $\displaystyle \int R(x)\frac{{\mathrm d}x}{\sqrt{W(x)}}$ is quite general.
Let $\displaystyle R(x)=\frac{P(x)}{Q(x)}$, where $P(x)$ and $Q(x)$ are both polynomials in $x$. $Q(x)$ can be factorized in the real domain as:
\begin{equation}\label{factorization}
Q(x)=A\prod^{I}_{i=1}(x-c_i)^{M_i}
\prod^{J}_{j=1}(x^2+d_j x+e_j)^{N_j},
\end{equation}
where $A$, $c_i$, $d_j$ and $e_j$ are real numbers; $I$, $J$, $M_i$ and $N_j$ are positive integers.
For all $j\ (j=1\cdots J)$, the discriminant of the quadratic polynomial $x^2+d_jx+e_j$ is negative.
So the partial fraction expansion, in which a rational fraction can be split into a sum of simple polynomials, of $R(x)$ is expressed as:
\begin{equation}\label{partial fraction}
R(x)
=\sum^{K}_{i=0}a_ix^i
+\frac{1}{A}\left(
\sum^{I}_{i=1}\sum^{M_i}_{j=1}\frac{b_{ij}}{(x-c_i)^j}
+\sum^{J}_{j=1}\sum^{N_j}_{l=1}\frac{m_{j}x+n_j}{(x^2+d_j x+e_j)^l}
\right),
\end{equation}
where $b_{ij}$, $m_j$ and $n_j$ are real numbers.
$\displaystyle \sum^{K}_{i=0}a_ix^i$ is the remainder of $P(x)$ divided by $Q(x)$. For instance, let $\displaystyle R(x)=\frac{3x^4-6x^3+4x^2-9x-3}{x^3-2x^2+x-2}$, then
\begin{align*}
R(x)=3x+\frac{2x+1}{x^2+1}+\frac{-1}{x-2}.
\end{align*}
Hence the original integral is decomposed into a sum of simpler integrals.
The next step is to write $W(x)$ into a form with no odd power terms of $x$.
Since $W(x)$ has no real roots, its factorization in the real domain can be expressed as:
\begin{align}
W\left(x\right)&=\left(f_1x^2+2g_1x+h_1\right)\left(f_2x^2+2g_2x+h_2\right),
\end{align}
An auxiliary equation of $\xi$ is established as:
\begin{align}
(f_1g_2-f_2g_1)\xi^2+(f_1h_2-f_2h_1)\xi+(g_1h_2-g_2h_1)=0.
\end{align}
The equation always has two real roots $\xi_1$ and $\xi_2$ ($\xi_1>\xi_2$).
Let $\displaystyle z=\frac{x-\xi_1}{x-\xi_2}$ as a new integral variable, we have
\begin{align}\label{method substitution}
\frac{\mathrm{d}x}{\sqrt{W(x)}}&=
\frac{\mathrm{d}z}{\sqrt{(u_1z^2+v_1)(u_2z^2+v_2)}},
\end{align} 
where
\begin{align}
\notag
&u_1=f_1\xi_2+g_1, && v_1=-f_1\xi_1-g_1,\\
&u_2=f_2\xi_2+g_2, && v_2=-f_2\xi_1-g_2.
\end{align}
By this substitution, the integral is expressed in a form closely resembling the standard elliptic integral. With some algebraic manipulation, the integral is finally written into a closed-form expression, only containing elementary functions and the three kinds of standard elliptic integrals defined as follows:
\begin{align}
K\left(\tau\right)
&=\int_{0}^{1}\frac{\mathrm{d}x}{\sqrt{1-x^{2}}\sqrt{1-\tau^{2}x^{2}}},
\\
E\left(\tau\right)
&=\int_{0}^{1}\frac{\sqrt{1-\tau^{2}x^{2}}}{\sqrt{1-x^{2}}}\mathrm{d}x,
\\
\varPi_1\left(\epsilon,\ \tau\right)
&=\int_{0}^{1}\frac{1}{1-\epsilon x^{2}}\frac{\mathrm{d}x}{\sqrt{1-x^{2}}\sqrt{1-\tau^{2}x^{2}}},
\end{align}
where $0<\tau<1$. 
$K\left(\tau\right)$, $E\left(\tau\right)$ and 
$\varPi_1\left(\epsilon,\ \tau\right)$ are called the elliptic integrals of the first, second and third kind, respectively, which can be calculated by commands ``ellipticK'', ``ellipticE'' and ``ellipticPi'' in Matlab, respectively. 
The generalized form the of elliptic integral of the third kind is defined as:
\begin{align}
\varPi_j\left(\epsilon,\ \tau\right)
&=\int_{0}^{1}\frac{1}{\left(1-\epsilon x^{2}\right)^j}\frac{\mathrm{d}x}{\sqrt{1-x^{2}}\sqrt{1-\tau^{2}x^{2}}},\quad j\geq 0.
\end{align}
By definition, $\varPi_0\left(\epsilon,\ \tau\right)=K\left(\tau\right)$. 
When $j>1$, $\varPi_j\left(\epsilon,\ \tau\right)$ can be calculated conveniently via the recursive formulae (Byrd and Morris, {\co 2013}):
\begin{align}
\varPi_2\left(\epsilon,\ \tau\right) =& \frac{1}{2\left(\epsilon-1\right)\left(\tau^2-\epsilon\right)}
\big[
\epsilon E\left(\tau\right)+\left(\tau^2-\epsilon\right)K\left(\tau\right)
+\left(2\epsilon\tau^2+2\epsilon-\epsilon^2-3\tau^2\right)\varPi_1\left(\epsilon, \ \tau\right)
\big],
\\
\notag
\varPi_{j+3}\left(\epsilon, \tau\right) =& 
\frac{1}{\left(2j+4\right)\left(1-\epsilon\right)\left(\tau^2-\epsilon\right)}
\big[
\left(2j+3\right)\left(\epsilon^2-2\epsilon\tau^2-2\epsilon+3\tau^2\right)\varPi_{j+2}\left(\epsilon,\ \tau\right)
\\&
+\left(2j+2\right)\left(\epsilon\tau^2+\epsilon-3\tau^2\right)\varPi_{j+1}\left(\epsilon, \ \tau\right)
+\left(2j+1\right)\tau^2 \varPi_j\left(\epsilon, \ \tau\right)
\big], \quad j\geq 0.
\end{align}
It should be pointed out that the method can not work when both the source and receiver are located at the free surface, because the substitution (\ref{method substitution}) no longer exists.
In other words, the derivation technique presented here, together with all formulae presented in the subsequent sections, is solely valid for Lamb's problem of second and third kinds.
\section{Refracted PS and SP waves}\label{Section reflected wave1}
\begin{figure}
\centering
\includegraphics[width=.6\textwidth]
{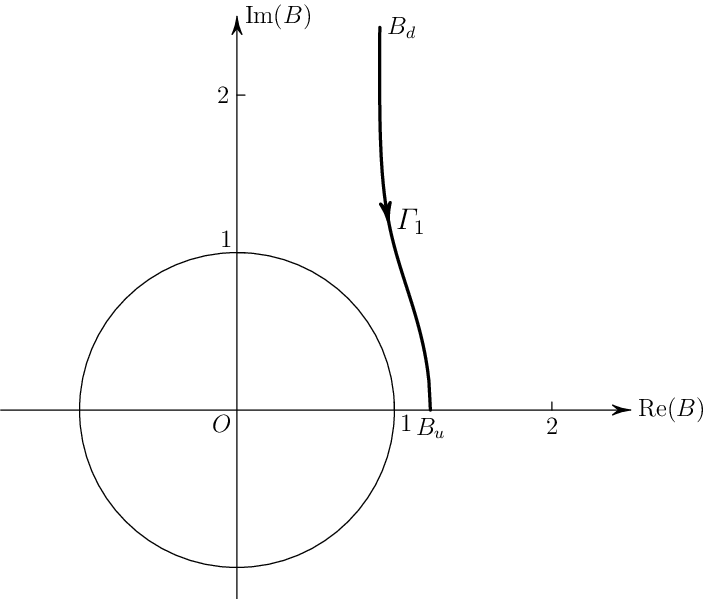}
\caption{Path of integration for variable $B$. 
The path from $B_d$ to $B_u$ is confined to the first quadrant outside the unit circle. 
$B_d$ is the root of eq. (\ref{QpB}) in the first quadrant with $p=0$, whereas $B_u$ is a real positive root of eq. (\ref{QpB}) with $p=p_u$.
}
\label{Figure path1}
\end{figure}

\begin{figure}
\centering
\includegraphics[width=.6\textwidth]
{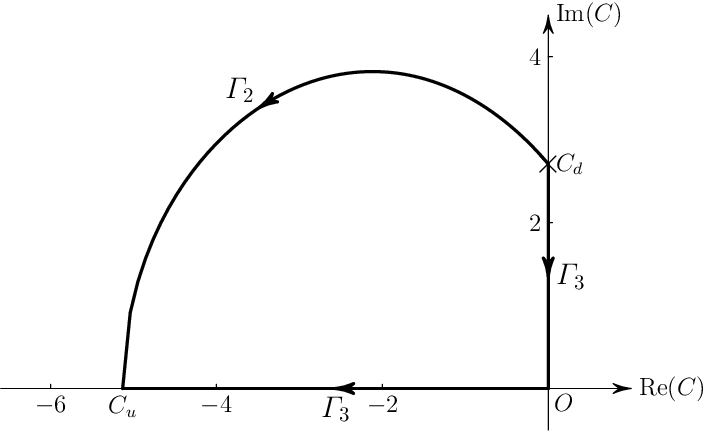}
\caption{Path of integration for variable $C$. The path is deformed from $\varGamma_2$ to $\varGamma_3$, and the endpoints $C_d$ and $C_u$ are fixed. The integral along the negative real axis has no contribution to the result. The branch point is denoted by the symbol $\times$.}
\label{Figure path2}
\end{figure}

In this section we derive the closed-from results ${\rm PS}$ and ${\rm SP}$ 
waves with the help of the theory on elliptic integrals introduced in the previous section.
The integral form of the reflected ${\rm PS}$ wave in eq. (\ref{PS Johnson}) is quite difficult to evaluate because of its irregular form. 
In particular, the procedure to obtain the upper integration limit $p_u$ is extremely complicated.
A substitution can be employed to change the integration in eq. (\ref{PS Johnson})
into a form more suitable for our purpose. 
Noticing that 
\begin{align}
\eta_{\beta}^2-\eta_{\alpha}^2=\beta^{-2}-\alpha^{-2}=\alpha^{-2}\kappa^2,
\end{align}
a new variable $B$ can be introduced to express $\eta_{\alpha}$ and $\eta_{\beta}$:
\begin{align}\label{B substitution}
\eta_{\alpha}=\kappa\frac{B^2-1}{2\alpha B},
\qquad
\eta_{\beta}=\kappa\frac{B^2+1}{2\alpha B}.
\end{align}
Given the requirements ${\rm Re}(\eta_\alpha)\geq 0$ and ${\rm Re}(\eta_\beta)\geq 0$, $B$ satisfies the conditions:
\begin{align}\label{B condition}
{\rm Re}(B) \geq 0, \qquad |B|\geq 1.
\end{align}
The integration variable $p$ in eq. (\ref{PS Johnson}) can then be expressed in terms of $B$:
\begin{align}\label{B and p}
p=\frac{\kappa}{2\alpha B}\sqrt{Q(B)},
\end{align}
where
\begin{align}
Q\left(B\right)=
a_4B^4+a_3B^3+a_2B^2+a_1B+a_0,
\end{align}
with
\begin{align}
\notag
&a_4 = \left(z+z'\right)^2+1,
&&a_3 = -\frac{4T_{\rm PS}}{\kappa}\left(z+z'\right),
\\
&a_2 = 2\left(z^2-z'^{2}+\frac{2T_{\rm PS}^2-2}{\kappa^2}\right),
&&a_1 = -\frac{4T_{\rm PS}}{\kappa}\left(z-z'\right),
\\
\notag
&a_0=\left(z-z'\right)^2+1.
\end{align}
Given a specific value of $p$ in $[0, p_u]$, $B$ is simply a root of the quartic polynomial:
\begin{align}\label{QpB}
Q(p, B)=a_4B^4+a_3B^3+\left(a_2-\frac{4\alpha^2 p^2}{\kappa^2}\right)B^2+a_1B+a_0.
\end{align}
As a quartic polynomial of $B$, $Q(p, B)$ has three fundamental properties:
\begin{enumerate}
\item $\forall p \in [0, p_u)$, eq. (\ref{QpB}) has two pairs of conjugate roots;
\item $\forall p \in [0, p_u]$, eq. (\ref{QpB}) has a pair of conjugate roots satisfying the condition (\ref{B condition});
\item When $p=p_u$, eq. (\ref{QpB}) has a pair of conjugate roots and a double real root;
\end{enumerate}
According to property (${\rm \rmnum{2}}$), when a specific value of $p$ is given, a pair of conjugate values for $B$ are allowed, which leads identical result due to the symmetry of the two integration paths. 
If the imaginary part of $B$ is limited to positive, the integral path for $B$ can be uniquely determined (Figure \ref{Figure path1}). 
It starts from $B_d$, which is the root of eq. (\ref{QpB}) in the first quadrant with $p=0$, i.e. $Q(0, B_d)=0$, and ends at $B_u$, a real positive root of eq. (\ref{QpB}) with $p=p_u$, i.e. $Q_p(p_u, B_u)=0$.
According to the properties $({\rm \rmnum{1}})$ and $({\rm \rmnum{3}})$, two simple formulae can be established to calculate $B_u$ and $p_u$:
\begin{align}
&2a_4B_u^4+a_3B_u^3-a_1B_u-2a_0=0, 
\\
&p_u=\frac{\alpha^{-1}\kappa}{2B_u}\left(a_4B_u^4+a_3B_u^3+a_2B_u^2+a_1B_u+a_0\right).
\end{align}
Based on the substitutions in eq. (\ref{B substitution}), the original integral can be written as:
\begin{align}
\ts{G}^{\rm PS}=-\frac{\kappa^2}{8R\pi^2\rho\alpha^2}
\mathcal{H}\left(t-t_{\rm PS}\right)
{\rm Im}\left\{\int^{B_u}_{B_d}
\frac{\ts{M}^{\rm PS}(B)}{B^3\sigma^{\rm PS}(B)}
\frac{{\rm d}B}{\sqrt{Q\left(B\right)}}\right\},
\end{align}
where
\begin{align}\label{GPS co1}
\ts{M}^{\rm PS}(B)=\left(B^4-1\right)\left(B^4-\frac{2}{\kappa^2}B^2+1\right)\ts{N}^{\rm PS}(B),
\end{align}
with
\begin{small}
\begin{align}\label{GPS co2}
\ts{N}^{\rm PS}(B)=\begin{bmatrix}
-\sin^2{\phi}I_1(B)+\cos{2\phi}I_2(B) & 
\sin{\phi}\cos{\phi}I_1(B)+\sin{2\phi}I_2(B) &
P(B)\left(B^4-1\right)\cos{\phi} \\
\sin{\phi}\cos{\phi}I_1(B)+\sin{2\phi}I_2(B) &
-\cos^2{\phi}I_1(B)-\cos{2\phi}I_2(B) &
P(B)\left(B^4-1\right)\sin{\phi} \\
-P(B)Z(B)\cos{\phi} &
-P(B)Z(B)\sin{\phi} &
-Z(B)\left(B^2-1\right)\\
\end{bmatrix},
\end{align}
\end{small}
and
\begin{align}\label{GPS co3}
\notag
&I_1(B)=P^2(B)\left(B^2+1\right),
&&I_2(B)=P^2(B)\left(B^2-1\right),
\\
&Z(B)=B^4-\left(\frac{4}{\kappa^2}+2\right)B^2+1,
&&P(B)=(z+z')B^2-\frac{2T_{\rm PS}}{\kappa}B+z-z'.
\end{align}
$\sigma^{\rm PS}(B)$ is a six-order polynomial of $B$:
\begin{align}\label{sigma_PS B}
\sigma^{\rm PS}(B)=B^6
+\left(\frac{2}{\kappa^4}+1\right)B^4
-\left(\frac{4}{\kappa^2}+1\right)B^2+1.
\end{align}
When the Poisson's ratio is less than 0.2613, which correspond to the P-to-S wave speed ratio $k$ in the range 1.4142-1.7636, $\sigma^{\rm PS}(B)$ has four real roots $\sigma^{\rm PS}_{m}$ ($m=1,\cdots,4$) and two purely imaginary roots
$\sigma^{\rm PS}_5$ and $\sigma^{\rm PS}_6$ ($\sigma^{\rm PS}_5=-\sigma^{\rm PS}_6$). $\sigma^{\rm PS}_5$ and $\sigma^{\rm PS}_6$ are associated with the emergence of the Rayleigh wave (FZ18), hence $\sigma^{\rm PS}(B)$ is called the \textit{Rayleigh function}, and $\sigma^{\rm PS}_5$ and $\sigma^{\rm PS}_6$ are called the \textit{Rayleigh zeros}. 
It is worth pointing out that the reflected PP and SS waves contain similar \textit{Rayleigh functions} $\sigma^{\rm PP}$ and $\sigma^{\rm SS}$, respectively, of which the purely imaginary roots are also associated with the Rayleigh wave. 
More features of the Rayleigh wave will be analyzed in Section \ref{Section Numerical Results}.
Noticing that the degree of every component of $\ts{M}^{\rm PS}(B)$ is eleven, $\frac{M_{ij}^{\rm PS}(B)}{B^3\sigma^{\rm PS}(B)}$ can be decomposed into:
\begin{align}\label{GPS coeffient}
\frac{M_{ij}^{\rm PS}(B)}{B^3\sigma^{\rm PS}(B)}
=\sum_{m=1}^{4}\frac{v^{\rm PS,\Rmnum{1}}_{m,ij}}{B-\sigma^{\rm PS}_{m}}
+\frac{v^{\rm PS,\Rmnum{2}}_{5,ij}+v^{\rm PS,\Rmnum{3}}_{6,ij}B}
{B^2+\left(\tilde{\sigma}_5^{\rm PS}\right)^2}
+\sum_{m=0}^{5}v^{\rm PS,\Rmnum{4}}_{ij,m}B^m
+\sum_{m=1}^{3}v^{\rm PS,\Rmnum{5}}_{ij,m}B^{-m},
\end{align}
where $\sigma_5^{\rm PS}={\rm i}\tilde{\sigma}_5^{\rm PS}$.
Hence $G^{\rm PS}_{ij}$ can be represented as a sum of simpler integrals as:
\begin{align}\label{GPS}
\notag
G^{\rm PS}_{ij}
=&
\frac{\kappa^2\mathcal{H}\left(t-t_{\rm PS}\right)}{8R\pi^2\rho\alpha^2}
\Big(
\sum^{4}_{m=1}v^{\rm PS,\Rmnum{1}}_{m,ij}V^{\rm PS,\Rmnum{1}}
\left(\sigma^{\rm PS}_{m}\right)
+v^{\rm PS,\Rmnum{2}}_{5,ij}V^{\rm PS,\Rmnum{2}}
\left(\tilde{\sigma}_5^{\rm PS},0\right)
+v^{\rm PS,\Rmnum{3}}_{5,ij}V^{\rm PS,\Rmnum{3}}
\left(\tilde{\sigma}_5^{\rm PS},0\right)
\\&
+\sum^{5}_{m=0}v^{\rm PS,\Rmnum{4}}_{m,ij}V^{\rm PS,\Rmnum{4}}_{m}
+\sum^{3}_{m=1}v^{\rm PS,\Rmnum{5}}_{m,ij}V^{\rm PS,\Rmnum{5}}_{m}
\Big)
,
\end{align}
where
\begin{align}
\notag
&V^{\rm PS,\Rmnum{1}}\left(a\right)
={\rm Im}\int^{B_u}_{B_d}
\frac{1}{B-a}\frac{\dd B}{\sqrt{Q(B)}},
&&
V_m^{\rm PS,\Rmnum{4}}
={\rm Im}\int^{B_u}_{B_d}
B^m\frac{\dd B}{\sqrt{Q(B)}},
\\
&
V^{\rm PS,\Rmnum{2}}\left(a,b\right)
={\rm Im}\int^{B_u}_{B_d}
\frac{1}{\left(B+b\right)^2+a^2}\frac{\dd B}{\sqrt{Q(B)}},
&&
V_m^{\rm PS,\Rmnum{5}}
={\rm Im}\int^{B_u}_{B_d}
B^{-m}\frac{\dd B}{\sqrt{Q(B)}},
\\
\notag
&
V^{\rm SS,\Rmnum{3}}\left(a,b\right)
={\rm Im}\int^{B_u}_{B_d}
\frac{B+b}{\left(B+b\right)^2+a^2}\frac{\dd B}{\sqrt{Q(B)}}.
\end{align}
Since $Q(B)$ has no real roots, it can be factored as:
\begin{align}
Q(B)=a_4\left(B^2+2q_1B+r_1\right)\left(B^2+2q_2B+r_2\right).
\end{align}
The discriminants of $B^2+2q_1B+r_1$ and $B^2+2q_2B+r_2$ are both negative.
As mentioned in Section 3, a new integral variable $\displaystyle C=\frac{B-\xi_1}{B-\xi_2}$ can be introduced in order to transform $Q(B)$ into a form with no odd power terms.
Two values $\xi_1$ and $\xi_2\ (\xi_1>\xi_2)$ satisfy the quadratic equation:
\begin{align}\label{xi PS}
(q_2-q_1)\xi^2+(r_2-r_1)\xi+q_1r_2-q_2r_1=0.
\end{align}
More attention should be paid to the deformation of the integral path by the substitution. 
The original path of integration $\varGamma_1$ (Figure \ref{Figure path1}) is changed to $\varGamma_2$ in the second quadrant 
(Figure \ref{Figure path2}).
The starting point $B_d$ is changed to $C_{d}$ at $(0, {\rm i}\sqrt{-\frac{\xi_1+q_1}{\xi_2+q_1}})$, the branch point of the integrand, and ending point $B_{u}$ changed to $C_{u}$ located on the negative real axis. Based on the Cauchy theorem, $\varGamma_2$ can be deformed into $\varGamma_3$, which goes first along the imaginary axis from $C_d$ to the origin, then follows the real axis and ends to $C_{u}$. When taking the imaginary part, the integral along the real axis has no contribution to the result.
After some algebraic manipulation, $V^{\rm PS,\Rmnum{1}}$, $V^{\rm PS,\Rmnum{2}}$ and $V^{\rm PS,\Rmnum{3}}$ can be written as:
\begin{align}
\label{PS V1}
V^{\rm PS,\Rmnum{1}}\left(a\right)=& \frac{1}{\xi_2-a}
M_{\rm PS}K\left(\tau_{_{\rm PS}}\right)
+\frac{\xi_2-\xi_1}{\left(\xi_1-a\right)\left(\xi_2-a\right)}
M_{\rm PS}\varPi_1\left(\frac{\left(\xi_2-a\right)^2}{\left(a-\xi_1\right)^2}\epsilon_{_{\rm PS}},\ \tau_{_{\rm PS}}\right),
\\ 
\label{PS V2}
V^{\rm PS,\Rmnum{2}}\left(a,b\right)=&
-\frac{\xi_1-\xi_2}{\upsilon_1^2\upsilon_2}M_{\rm PS}
{\rm Re}\left(m_{_{\rm PS}}\varPi_1\left(q_{_{\rm PS}},\ \tau_{_{\rm PS}}\right)\right)
+\frac{M_{\rm PS}}{\upsilon_2}
K\left(\tau_{_{\rm PS}}\right)
,\\ 
\label{PS V3}
V^{\rm PS,\Rmnum{3}}\left(a,b\right)=&
\frac{\xi_1-\xi_2}{\upsilon_1^2\upsilon_2}M_{\rm PS}
{\rm Re}\left(n_{_{\rm PS}}\varPi_1\left(q_{_{\rm PS}},\ \tau_{_{\rm PS}}\right)\right)
+\frac{M_{\rm PS}}{\upsilon_2}\left(\xi_2+b\right)
K\left(\tau_{_{\rm PS}}\right),
\end{align}
where
\begin{align}\label{PS parameter1}
&
\epsilon_{_{\rm PS}}=\frac{\xi_1+q_1}{\xi_2+q_1},
&&
\tau_{_{\rm PS}}=\sqrt{\frac{(\xi_2+q_2)(\xi_1+q_1)}{(\xi_2+q_1)(\xi_1+q_2)}},
&&
M_{\rm PS}=\frac{1}{\sqrt{-\left(\left(z+z'\right)^2+1\right)
\left(\xi_2+q_1\right)\left(\xi_1+q_2\right)}},
\end{align}
and
\begin{align}\label{PS parameter2}
\notag
q_{_{\rm PS}}=&-\frac{\epsilon_{_{\rm PS}}}{\upsilon_1^2}
\Big[
a^2\left(\xi_1-\xi_2\right)^2
-\upsilon_3^2
+2{\rm i}a\upsilon_3\left(\xi_1-\xi_2\right)
\Big],
\\
m_{_{\rm PS}}=&
\upsilon_0\upsilon_1-
\frac{{\rm i}}{{\rm Im}\left(q_{_{\rm PS}}\right)}
\Big[
\epsilon_{_{\rm PS}}
\left(\upsilon_0\upsilon_2
-4\left(\xi_2+b\right)\upsilon_3\right)
+{\rm Re}\left(m_{_{\rm PS}}\right){\rm Re}\left(q_{_{\rm PS}}\right)
\Big],
\\
\notag
n_{_{\rm PS}}=&
\upsilon_1\upsilon_4-
\frac{{\rm i}}{{\rm Im}\left(q_{_{\rm PS}}\right)}
\Big[
\epsilon_{_{\rm PS}}
\Big(
2\left(a^2-\left(\xi_2+b\right)^2\right)\upsilon_3
-\upsilon_2\upsilon_4
\big)
+{\rm Re}\left(n_{_{\rm PS}}\right){\rm Re}\left(q_{_{\rm PS}}\right)
\Big],
\end{align}
with
\begin{align}\label{PS parameter3}
&\upsilon_0=\xi_1+\xi_2+2b, 
&&\upsilon_1=a^2+\left(\xi_1+b\right)^2,
&&\upsilon_2=a^2+\left(\xi_2+b\right)^2,
\\
\notag
&\upsilon_3=a^2+\left(\xi_1+b\right)\left(\xi_2+b\right),
&&\upsilon_4=a^2-\left(\xi_1+b\right)\left(\xi_2+b\right),
\end{align}
Both $V_m^{\rm PS,\Rmnum{4}}$ and $V_m^{\rm PS,\Rmnum{5}}$ can be expressed as products between $M_{\rm PS}$ and linear sums of elliptic integrals $\varPi_j$ as follows:
\begin{align}
\label{PS V4}
V_m^{\rm PS,\Rmnum{4}}=&
M_{\rm PS}\sum^{m}_{j=0}b_{mj}\varPi_j
\left(\epsilon_{_{\rm PS}},\tau_{_{\rm PS}}\right), 
\quad m=0,\cdots,5
\\
\label{PS V5}
V_m^{\rm PS,\Rmnum{5}}=&
M_{\rm PS}\sum^{m}_{j=1}d_{mj}\varPi_j
\left(\epsilon_{_{\rm PS}}\xi_2^2/\xi_1^2,\tau_{_{\rm PS}}\right), 
\quad m=1,\cdots,3
\end{align}
where
\begin{align}\label{PS parameter4}
\notag
&b_{00} = 1,
&&b_{10} =\xi_2,
\\
\notag
&b_{11}=\xi_1-\xi_2,
&&b_{20}=\xi_2^2,
\\
\notag
&b_{21} = \left(\xi_1-\xi_2\right)
\left(3\xi_2-\xi_1\right),
&&b_{22} = 2\left(\xi_1-\xi_2\right)^2,
\\
\notag
&b_{30} = \xi_2^3,
&&b_{31} = 3\xi_2\left(\xi_1-\xi_2\right)
\left(2\xi_2-\xi_1\right),
\\
&b_{32} = 3\xi_2\left(\xi_1-\xi_2\right)^2
\left(3\xi_2-\xi_1\right),
&&b_{33} = 4\left(\xi_1-\xi_2\right)^3,
\\
\notag
&
b_{40} = \xi_2^4,
&&
b_{41} = 2\xi_2^2\left(\xi_1-\xi_2\right)
\left(5\xi_2-3\xi_1\right),
\\
\notag
&
b_{42} = \left(\xi_1-\xi_2\right)^2
\left(\xi^2_1-14\xi_1\xi_2+25\xi_2^2\right),
&&
b_{43} = 8\left(\xi_1-\xi_2\right)^3
\left(3\xi_2-\xi_1\right),
\\
\notag
&
b_{44} = 8\left(\xi_1-\xi_2\right)^4,
&&
b_{50} = \xi_2^5,
\\
\notag
&
b_{51} = 5\xi_2^3\left(\xi_1-\xi_2\right)
\left(3\xi_1-2\xi_2\right),
&&
b_{52} = 5\xi_2\left(\xi_1-\xi_2\right)^2
\left(\xi_1^2-8\xi_1\xi_2+11\xi_2^2\right),
\\
\notag
&
b_{53} = 5\left(\xi_1-\xi_2\right)^3
\left(\xi_1^2-10\xi_1\xi_2+17\xi_2^2\right),
&&
b_{54} = 20\left(\xi_1-\xi_2\right)^4\left(3\xi_2-\xi_1\right),
\\
\notag
&
b_{55} = 16\left(\xi_1-\xi_2\right)^5,
\end{align}
and
\begin{align}\label{PS parameter5}
\notag
&
d_{10} = \frac{1}{\xi_2},
&&
d_{11} = \frac{\xi_2-\xi_1}{\xi_1\xi_2},
\\
&
d_{20} = \frac{1}{\xi_2^2},
&&
d_{21} = \frac{2}{\xi_2}\frac{\xi_2-\xi_1}{\xi_1\xi_2}
-\frac{\left(\xi_2-\xi_1\right)^2}{\xi_1^2\xi_2^2},
\\
\notag
&
d_{22} = 2\frac{\left(\xi_2-\xi_1\right)^2}{\xi_1^2\xi_2^2},
&&
d_{30} = \frac{1}{\xi_2^3},
\\
\notag
&
d_{31} = \frac{3}{\xi_2^2}\frac{\xi_2-\xi_1}{\xi_1\xi_2}
-\frac{3}{\xi_2}\frac{\left(\xi_2-\xi_1\right)^2}{\xi_1^2\xi_2^2},
&&
d_{32} = \frac{6}{\xi_2}\frac{\left(\xi_2-\xi_1\right)^2}{\xi_1^2\xi_2^2}
-3\frac{\left(\xi_2-\xi_1\right)^3}{\xi_1^3\xi_2^3},
\\
\notag
&
d_{33} = 4\frac{\left(\xi_2-\xi_1\right)^3}{\xi_1^3\xi_2^3}.
\end{align}
The expressions for $\ts{G}^{\rm SP}$ can be obtained from those for 
$\ts{G}^{\rm PS}$ by virtual of the reciprocity of the Green's functions:
\begin{align}\label{GSP}
\ts{G}^{\rm SP}=\begin{bmatrix}
G^{\rm PS\star}_{11} & 
G^{\rm PS\star}_{12} &
-G^{\rm PS\star}_{31} \\
G^{\rm PS\star}_{21} &
G^{\rm PS\star}_{22} &
-G^{\rm PS\star}_{32} \\
-G^{\rm PS\star}_{13} &
-G^{\rm PS\star}_{23} &
G^{\rm PS\star}_{33} \\
\end{bmatrix},
\end{align}
where the symbol $\star$ indicates a switch when the positions of $x'_3$ and $x_3$ are interchanged for the Green's function. 
\section{Reflected PP and SS waves and surface refracted S-S wave}\label{Section reflected wave2}
In this section, the final results of $\ts{G}^{\rm PP}$, $\ts{G}^{\rm SS}$ and $\ts{G}^{\rm S\text{-}S}$ are presented, respectively. 
More details about the derivations can be found in FZ18. 
Throughout this section, $r$ and $\theta$ are redefined to account for the reflected ray paths:
\begin{align}
r=\sqrt{R^2+\left(x'_3+x_3\right)^2},
\quad
\theta=\text{arctan}\left(\frac{R}{x'_3+x_3}\right),
\end{align} 
and $T_{\rm dP}$ and $T_{\rm dS}$ are substituted by $T_{\rm PP}$ and $T_{\rm SS}$ for the dimensionless times in $\ts{G}^{\rm PP}$ and $\ts{G}^{\rm SS}$, respectively.
\subsection{Reflected PP wave}
The expressions for the components of the reflected PP wave can be written as:
\begin{align}\label{GPP}
\notag
G^{\rm PP}_{ij}=&\frac{\mathcal{H}\left(T_{\rm PP}-1\right)}{\pi^{2}\mu r}
\Big\{
u^{\rm PP}_{ij,1}U^{\rm PP}_{1}\left(\sigma_1^{\rm PP}\right)
+u^{\rm PP}_{ij,2}U^{\rm PP}_{1}\left(\sigma_2^{\rm PP}\right)
+u^{\rm PP}_{ij,3}U^{\rm PP}_{1}\left(\sigma_3^{\rm PP}\right)
+u^{\rm PP}_{ij,4}U^{\rm PP}_{1}\left(\sigma_4^{\rm PP}\right)
\\
\notag
&
+u^{\rm PP}_{ij,5}U^{\rm PP}_{2}\left(\tilde{\sigma}_5^{\rm PP}\right)
+u^{\rm PP}_{ij,6}U^{\rm PP}_{3}\left(\tilde{\sigma}_6^{\rm PP}\right)
+u^{\rm PP}_{ij,7}U^{\rm PP}_{4}
+u^{\rm PP}_{ij,8}U^{\rm PP}_{5}
+u^{\rm PP}_{ij,9}U^{\rm PP}_{6}
\\
\notag
&+u^{\rm PP}_{ij,10}U^{\rm PP}_{7}
+u^{\rm PP}_{ij,11}U^{\rm PP}_{8}
+v^{\rm PP}_{ij,1}V^{\rm PP}_{1}\left(\sigma_1^{\rm PP}\right)
+v^{\rm PP}_{ij,2}V^{\rm PP}_{1}\left(\sigma_2^{\rm PP}\right)
+v^{\rm PP}_{ij,3}V^{\rm PP}_{1}\left(\sigma_3^{\rm PP}\right)
\\ \notag
&
+v^{\rm PP}_{ij,4}V^{\rm PP}_{1}\left(\sigma_4^{\rm PP}\right)
+v^{\rm PP}_{ij,5}V^{\rm PP}_{2}\left(\tilde{\sigma}_5^{\rm PP}\right)
+v^{\rm PP}_{ij,6}V^{\rm PP}_{3}\left(\tilde{\sigma}_6^{\rm PP}\right)
+v^{\rm PP}_{ij,7}V^{\rm PP}_{4}
+v^{\rm PP}_{ij,8}V^{\rm PP}_{5}
\\&
+v^{\rm PP}_{ij,9}V^{\rm PP}_{6}
+v^{\rm PP}_{ij,10}V^{\rm PP}_{7}
+v^{\rm PP}_{ij,11}V^{\rm PP}_{8}
+v^{\rm PP}_{ij,12}V^{\rm PP}_{9}
\Big\},
\end{align}
where $\sigma_5^{\rm PP}={\rm i}\tilde{\sigma}_5^{\rm PP}$ and 
$\sigma_6^{\rm PP}={\rm i}\tilde{\sigma}_6^{\rm PP}$. 
$u_{ij,m}^{\rm PP}\ (m=1,\cdots,11)$, $v_{ij,m}^{\rm PP}\ (m=1,\cdots,12)$ and $\sigma_m^{\rm PP}\ (m=1,\cdots,6)$ denote the coefficients determined by the partial fraction expansion, which are given in the Appendix \ref{Section coefficient}, and
$U_n^{\rm PP} \ (n=1,\cdots,8)$ and $V_n^{\rm PP} \ (n=1,\cdots,9)$ denote the expressions containing elementary functions as well as elliptic integrals.
The expressions of $U^{\rm PP}_{n} \ (n=1,\cdots,6)$ are identical to those in eqs. (29)--(31) in FZ18, and $V^{\rm PP}_{n} \ (n=1,\cdots,7)$ are identical to those in eqs. (33)--(39) in FZ18. $U_7^{\rm PP}$, $U_8^{\rm PP}$, $V_8^{\rm PP}$ and $V_9^{\rm PP}$ are expressed as follows:
\begin{align}
\label{PP U7}
U_7^{\rm PP}
=&\frac{\pi}{2}T_{\rm PP}^{3}\cos^{3}{\theta}
-\frac{3\pi}{4}T_{\rm PP}\cos{\theta}\left(T_{\rm PP}^{2}-1\right)\sin^{2}{\theta},
\\
\label{PP U8}
U_8^{\rm PP}
=&\frac{\pi}{2}T_{\rm PP}^{4}\cos^{4}{\theta}
-\frac{3\pi}{2}T^2_{\rm PP}\cos^2{\theta}\left(T_{\rm PP}^{2}-1\right)\sin^{2}{\theta}
+\frac{3\pi}{16}\left(T_{\rm PP}^{2}-1\right)^2\sin^{4}{\theta},
\\
\label{PP V8}
V_8^{\rm PP}=&
M_{\rm PP}\sum^{4}_{j=0}b_{4j}\varPi_j
\left(\epsilon_{_{\rm PP}},\tau_{_{\rm PP}}\right),
\\
\label{PP V9}
V_9^{\rm PP}=&
M_{\rm PP}\sum^{5}_{j=0}b_{5j}\varPi_j
\left(\epsilon_{_{\rm PP}},\tau_{_{\rm PP}}\right),
\end{align}
where
\begin{align}\label{PP parameter}
&
M_{\rm PP}=\frac{1}{\sqrt{\xi_{1}(T_{\rm PP}\cos{\theta}-\xi_{2})}}, 
&&
\epsilon_{\rm PP}=\frac{T_{\rm PP}\cos\theta-\xi_1}{T_{\rm PP}\cos\theta-\xi_2},
&& 
\tau_{_{\rm PP}}=\sqrt{\frac{\epsilon_{_{\rm PP}}\xi_2}{\xi_1}}, 
\end{align}
where $\xi_1$ and $\xi_2$ ($\xi_1>\xi_2$) are the two roots of the equation:
\begin{align}\label{xi PP}
\xi^{2}-\frac{T_{\rm PP}^{2}+\cos^{2}{\theta}-k^{2}}{T_{\rm PP}\cos{\theta}}\xi+1-k^{2}=0.
\end{align}
\subsection{Reflected SS wave}
The expressions for the components of the reflected SS wave can be written as:
\begin{align}\label{GSS}
\notag
G^{\rm SS}_{ij}=&\frac{\mathcal{H}\left(T_{\rm SS}-1\right)}{\pi^{2}\mu r}
\Big\{
u^{\rm SS}_{ij,1}U^{\rm SS}_{1}\left(\sigma_1^{\rm SS}\right)
+u^{\rm SS}_{ij,2}U^{\rm SS}_{1}\left(\sigma_2^{\rm SS}\right)
+u^{\rm SS}_{ij,3}U^{\rm SS}_{1}\left(\sigma_3^{\rm SS}\right)
+u^{\rm SS}_{ij,4}U^{\rm SS}_{1}\left(\sigma_4^{\rm SS}\right)
\\
\notag
&
+u^{\rm SS}_{ij,5}U^{\rm SS}_{2}\left(\tilde{\sigma}_5^{\rm SS}\right)
+u^{\rm SS}_{ij,6}U^{\rm SS}_{3}\left(\tilde{\sigma}_6^{\rm SS}\right)
+u^{\rm SS}_{ij,7}U^{\rm SS}_{4}
+u^{\rm SS}_{ij,8}U^{\rm SS}_{5}
+u^{\rm SS}_{ij,9}U^{\rm SS}_{6}
\\
\notag
&+u^{\rm SS}_{ij,10}U^{\rm SS}_{7}
+u^{\rm SS}_{ij,11}U^{\rm SS}_{8}
+v^{\rm SS}_{ij,1}V^{\rm SS}_{1}\left(\sigma_1^{\rm SS}\right)
+v^{\rm SS}_{ij,2}V^{\rm SS}_{1}\left(\sigma_2^{\rm SS}\right)
+v^{\rm SS}_{ij,3}V^{\rm SS}_{1}\left(\sigma_3^{\rm SS}\right)
\\ \notag
&
+v^{\rm SS}_{ij,4}V^{\rm SS}_{1}\left(\sigma_4^{\rm SS}\right)
+v^{\rm SS}_{ij,5}V^{\rm SS}_{2}\left(\tilde{\sigma}_5^{\rm SS}\right)
+v^{\rm SS}_{ij,6}V^{\rm SS}_{3}\left(\tilde{\sigma}_6^{\rm SS}\right)
+v^{\rm SS}_{ij,7}V^{\rm SS}_{4}
+v^{\rm SS}_{ij,8}V^{\rm SS}_{5}
\\
&
+v^{\rm SS}_{ij,9}V^{\rm SS}_{6}
+v^{\rm SS}_{ij,10}V^{\rm SS}_{7}
+v^{\rm SS}_{ij,11}V^{\rm SS}_{8}
+v^{\rm SS}_{ij,12}V^{\rm SS}_{9}
\Big\}.
\end{align}
where $\sigma_5^{\rm SS}={\rm i}\tilde{\sigma}_5^{\rm SS}$ and 
$\sigma_6^{\rm SS}={\rm i}\tilde{\sigma}_6^{\rm SS}$.
Similarly, all the coefficients $u_{ij,m}^{\rm SS}\ (m=1,\cdots,11)$, $v_{ij,m}^{\rm SS}\ (m=1,\cdots,12)$ and $\sigma_m^{\rm SS}\ (m=1,\cdots,6)$ are given in the Appendix \ref{Section coefficient}.
All expressions for $U_n^{\rm SS}\ (n=1,\cdots,8)$ are identical to the respective $U^{\rm PP}_{n}$ except for changing $T_{\rm PP}$ to $T_{\rm SS}$, and $V^{\rm SS}_{n} \ (n=1,\cdots,7)$ are identical to those in eqs. (61)--(67) in FZ18. $V^{\rm SS}_{8}$ and $V^{\rm SS}_{9}$ are expressed as follows:
\begin{align}
\label{SS V8}
V_8^{\rm SS}=&
M_{\rm SS}\sum^{4}_{j=1}D^j b_{4j}\varPi_j
\left(\epsilon_{_{\rm SS}},\tau_{_{\rm SS}}\right),
\\
\label{SS V9}
V_9^{\rm SS}=&
M_{\rm SS}\sum^{5}_{j=1}D^j b_{5j}\varPi_j
\left(\epsilon_{_{\rm SS}},\tau_{_{\rm SS}}\right),
\end{align}
where
\begin{align}\label{SS parameter}
&D=\frac{T_{\rm SS}\cos\theta-\xi_2}{\xi_1-\xi_2},
&&
M_{\rm SS}=\frac{1}{\sqrt{T_{\rm SS}\cos{\theta}(\xi_{1}-\xi_{2})}}, 
\\
&\epsilon_{_{\rm SS}}=\frac{\xi_1-T_{\rm SS}\cos\theta}{\xi_1-\xi_2},
&& 
\tau_{_{\rm SS}}=\sqrt{\frac{\xi_{2}(\xi_{1}-T_{\rm SS}\cos{\theta})}{T_{\rm SS}\cos{\theta}(\xi_{1}-\xi_{2})}}.
\end{align}
where $\xi_1$ and $\xi_2$ ($\xi_1>\xi_2$) are the two roots of the equation:
\begin{align}\label{xi SS}
\xi^{2}-\frac{T_{\rm SS}^{2}+\cos^{2}{\theta}-k^{-2}}{T_{\rm SS}\cos{\theta}}\xi+1-k^{-2}=0.
\end{align}
\subsection{Surface refracted S-S wave}
The expressions for the components of the surface refracted S-S wave can be written as:
\begin{align}\label{GS2S}
\notag
G^{\rm S\text{-}S}_{ij}&=
\frac{\mathcal{H}\left(k\sin\theta-1\right)}{\pi^{2}\mu r}
\left(\mathcal{H}\left(T_{\rm S\text{-}S}-1\right)-\mathcal{H}\left(T_{\rm SS}-1\right)\right)
\Big\{
v^{\rm SS}_{ij,1}V^{\rm S\text{-}S}_{1}\left(\sigma_1^{\rm SS}\right)
+v^{\rm SS}_{ij,2}V^{\rm S\text{-}S}_{1}\left(\sigma_2^{\rm SS}\right)
\\
\notag
&
+v^{\rm SS}_{ij,3}V^{\rm S\text{-}S}_{1}\left(\sigma_3^{\rm SS}\right)
+v^{\rm SS}_{ij,4}V^{\rm S\text{-}S}_{1}\left(\sigma_4^{\rm SS}\right)
+v^{\rm SS}_{ij,5}V^{\rm S\text{-}S}_{2}\left(\tilde{\sigma}_5^{\rm SS}\right)
+v^{\rm SS}_{ij,6}V^{\rm S\text{-}S}_{3}\left(\tilde{\sigma}_6^{\rm SS}\right)
\\
&
+v^{\rm SS}_{ij,7}V^{\rm S\text{-}S}_{4}
+v^{\rm SS}_{ij,8}V^{\rm S\text{-}S}_{5}
+v^{\rm SS}_{ij,9}V^{\rm S\text{-}S}_{6}
+v^{\rm SS}_{ij,10}V^{\rm S\text{-}S}_{7}
+v^{\rm SS}_{ij,11}V^{\rm S\text{-}S}_{8}
+v^{\rm SS}_{ij,12}V^{\rm S\text{-}S}_{9}
\Big\}.
\end{align}
$V^{\rm S\text{-}S}_{n} \ (n=1,\cdots,7)$ are identical to those in eqs. (71)--(77) in FZ18. $V^{\rm S\text{-}S}_{8}$ and $V^{\rm S\text{-}S}_{9}$ are expressed as follows:
\begin{align}
\label{S2S V8}
V_8^{\rm S\text{-}S}=&
M_{\rm S\text{-}S}\sum^{4}_{j=0}D^j b_{4j}\varPi_j
\left(\epsilon_{_{\rm S\text{-}S}},\tau_{_{\rm S\text{-}S}}\right)
+\sum^{m}_{j=1}c_{4j}U_j^{\rm S\text{-}S}, 
\\
\label{S2S V9}
V_9^{\rm S\text{-}S}=&
M_{\rm S\text{-}S}\sum^{5}_{j=0}D^j b_{5j}\varPi_j
\left(\epsilon_{_{\rm S\text{-}S}},\tau_{_{\rm S\text{-}S}}\right)
+\sum^{m}_{j=1}c_{5j}U_j^{\rm S\text{-}S},
\end{align}
where
\begin{align}
\label{S2S parameter}
&
U_1^{\rm S\text{-}S} = \frac{\pi}{2},
&&
U_2^{\rm S\text{-}S} = -\frac{\pi}{4}F,
\\
&
U_3^{\rm S\text{-}S} = \frac{\pi}{16}
\left(2F^2+T_{\rm SS}^2\cos^2\theta\right),
&&
U_4^{\rm S\text{-}S} = -\frac{\pi}{32}F
\left(2F^2+3T^2_{\rm SS}\cos^2\theta\right),
\\
&
U_5^{\rm S\text{-}S} = \frac{\pi}{32}
\left(8F^4+24F^2T^2_{\rm SS}\cos^2\theta+3T^4_{\rm SS}\cos^4\theta\right),
\end{align}
and
\begin{align}
&
F=2\xi_2-T_{\rm SS}\cos\theta,
&&
M_{\rm S\text{-}S}=
\frac{1}{\sqrt{\xi_{2}\left(\xi_{1}-T_{\rm SS}\cos{\theta}\right)}}, 
\\
&
\epsilon_{_{\rm S\text{-}S}}=\frac{T_{\rm SS}\cos\theta}{\xi_2},
&&
\tau_{_{\rm S\text{-}S}}=\sqrt{\frac{T_{\rm SS}\cos{\theta}\left(\xi_{1}-\xi_{2}\right)}{\xi_{2}\left(\xi_{1}-T_{\rm SS}\cos{\theta}\right)}},
\end{align}
\begin{align}
\notag
&
c_{41} = 4\xi_2^3,
&&
c_{42} = 4\xi_2\left(4\xi_2-\xi_1\right),
&&           
c_{43} = 4\left(5\xi_2-\xi_1\right),
\\
&              
c_{44} = 8,
&&
c_{51} = 5\xi_2^4,
&&
c_{52} = 10\xi_2^2\left(3\xi_2-\xi_1\right),
\\
\notag
&                      
c_{53} = 61\xi_2^2-22\xi_2\xi_1+\xi_1^2,
&&    
c_{54} = 4\left(13\xi_2-3\xi_1\right),
&&                          
c_{55} = 2.                
\end{align} 
$\xi_1$ and $\xi_2$ ($\xi_1>\xi_2$) are the two roots of the equation:
\begin{align}\label{xi S2S}
\xi^{2}-\frac{T_{\rm SS}^{2}+\cos^{2}{\theta}-k^{-2}}{T_{\rm SS}\cos{\theta}}\xi+1-k^{-2}=0.
\end{align}
\section{Ultimate results for $\ts{G}$}\label{Section guide}
So far, the complete closed-form formulae for $\ts{G}$ have been obtained. However, the solution scheme is sufficiently complicated that readers may not know what the ultimate formulae are.
An explicit guide should be listed to help readers to implement our results easily: 
\begin{enumerate}
\item $\ts{G}$ is calculated by (\ref{Gfull}), which consists of seven parts: 
$\ts{G}^{\rm dP}$, $\ts{G}^{\rm dS}$, $\ts{G}^{\rm PP}$, 
$\ts{G}^{\rm SS}$, $\ts{G}^{\rm PS}$, $\ts{G}^{\rm SP}
$ and $\ts{G}^{\rm S\text{-}S}$.
\item $\ts{G}^{\rm dP}$ is calculated by (\ref{GdP}) and (\ref{GdP component}).
\item $\ts{G}^{\rm dS}$ is calculated by (\ref{GdS}) and (\ref{GdS component}).
\item $\ts{G}^{\rm PS}$ is calculated by (\ref{GPS}), which consists of the functions labeled as $V^{\rm PS}$, and the coefficients labeled as $v^{\rm PS}$ and $\sigma^{\rm PS}$:
\begin{enumerate}
\item The functions in (\ref{GPS}) are calculated by (\ref{PS V1}), (\ref{PS V2}), (\ref{PS V3}), (\ref{PS V4}) and (\ref{PS V5}),
and the related parameters are calculated by (\ref{xi PS}), (\ref{PS parameter1}), (\ref{PS parameter2}), (\ref{PS parameter3}), (\ref{PS parameter4}) and (\ref{PS parameter5}).
\item The coefficients in (\ref{GPS}) are calculated by (\ref{GPS coeffient}), and the related terms are in (\ref{GPS co1})--(\ref{sigma_PS B}).
\end{enumerate}
\item $\ts{G}^{\rm SP}$ is calculated by (\ref{GSP}).
\item $\ts{G}^{\rm PP}$ is calculated by (\ref{GPP}), which consists of the functions labeled as $U^{\rm PP}$ and $V^{\rm PP}$, and the coefficients labeled as $u^{\rm PP}$, $v^{\rm PP}$ and $\sigma^{\rm PP}$:
\begin{enumerate}
\item The functions in (\ref{GPP}) are calculated by (\ref{PP U7})--(\ref{PP V9}), and eqs. (29)--(31), (33)--(39) in FZ18.
The related parameters are calculated by (\ref{PP parameter}) and (\ref{xi PP}).
\item The coefficients in (\ref{GPP}) are calculated by (\ref{GPP coeffient1}) and (\ref{GPP coeffient2}), and the related terms are in (\ref{GPP co}).
\end{enumerate}
\item $\ts{G}^{\rm SS}$ is calculated by (\ref{GSS}), which consists of the functions labeled as $U^{\rm SS}$ and $V^{\rm SS}$, and the coefficients labeled as $u^{\rm SS}$, $v^{\rm SS}$ and $\sigma^{\rm SS}$:
\begin{enumerate}
\item The functions in (\ref{GSS}) are calculated by (\ref{SS V8})--(\ref{SS V9}), and eqs. (61)--(67) in FZ18.
Other parameters are calculated by (\ref{SS parameter}) and (\ref{xi SS}).
\item The coefficients in (\ref{GSS}) are calculated by (\ref{GSS coeffient1}) and (\ref{GSS coeffient2}), the related terms are in (\ref{GSS co1})--(\ref{GSS co3}).
\end{enumerate}
\item $\ts{G}^{\rm S\text{-}S}$ is calculated by (\ref{GS2S}), which consists of the functions labeled as $V^{\rm S\text{-}S}$, and the coefficients labeled as $u^{\rm SS}$, $v^{\rm SS}$ and $\sigma^{\rm SS}$:
\begin{enumerate}
\item The functions in (\ref{GS2S}) are calculated by (\ref{S2S V8})--(\ref{S2S V9}), and eqs. (71)--(77) in FZ18.
Other parameters are calculated by (\ref{S2S parameter}) and (\ref{xi S2S}).
\item The coefficients in (\ref{GS2S}) are the same to those in $\ts{G}^{\rm SS}$.
\end{enumerate}
\end{enumerate}
Furthermore, a Matlab program for implementing our solution will be provided as Supporting Information on the journel site, to allow comparisons with other methods.
\section{Numerical Results}\label{Section Numerical Results}
In this section, we present a few numerical examples of the Green's functions and compare them with the results of J74 to validate our formulae.
Furthermore, we also extract the terms related to the emergence of the Rayleigh wave 
from our closed-form solutions.
Those terms are known as the \textit{Rayleigh terms}, and we demonstrate that numerical results from the Rayleigh terms contain all the characteristics of the Rayleigh wave.

In all the numerical calculations shown here, the material parameters are set as: $\alpha=8.00\ \rm{km/s}$, $\beta=4.62\ {\rm km/s}$, and $\rho=3.30\ {\rm g/cm^3}$, the same as those in J74. 
\begin{figure}
\centering
\subfigure{
\begin{minipage}{.48\textwidth}
\centering
\includegraphics[scale=0.6]
{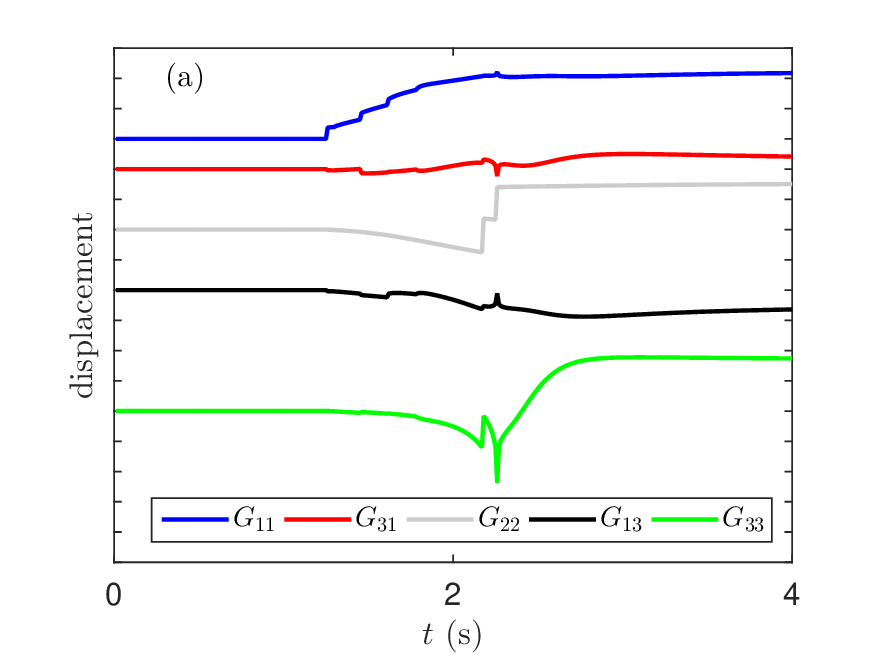}
\end{minipage}
}
\subfigure{
\begin{minipage}{.48\textwidth}
\centering
\includegraphics[scale=0.6]
{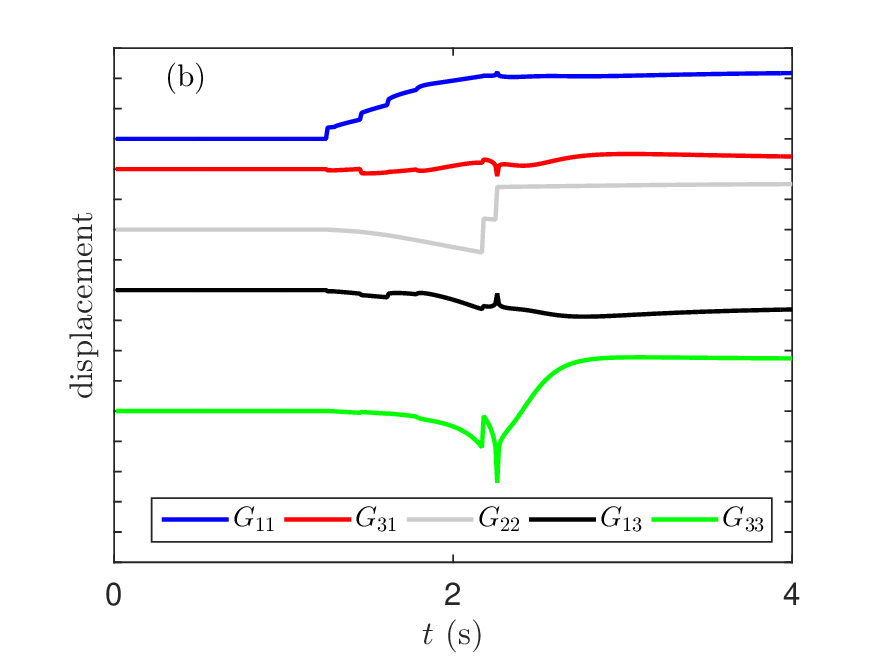}
\end{minipage}
}
\caption{Comparison of the time series of $\ts{G}$ between (a) our results and (b) those in J74. The coordinates of the source and the receiver are ($0$, $0$, $2$ km) and ($10$ km, $0$, $1$ km), respectively. 
Only the non-zero components are shown.
For a force of 1 Newton with Heaviside step function time history, 
each coordinate interval on the vertical axis is equal to $10^{-16}$ m.  
}
\label{Figure compare johnson}
\end{figure} 
Figure \ref{Figure compare johnson} compares the time series of the non-zero components of the tensor Green's function between this study and J74. 

The two sets of solutions agree perfectly and completely, which conclusively validates our results.
\begin{figure}
\centering
\includegraphics[scale=0.7]
{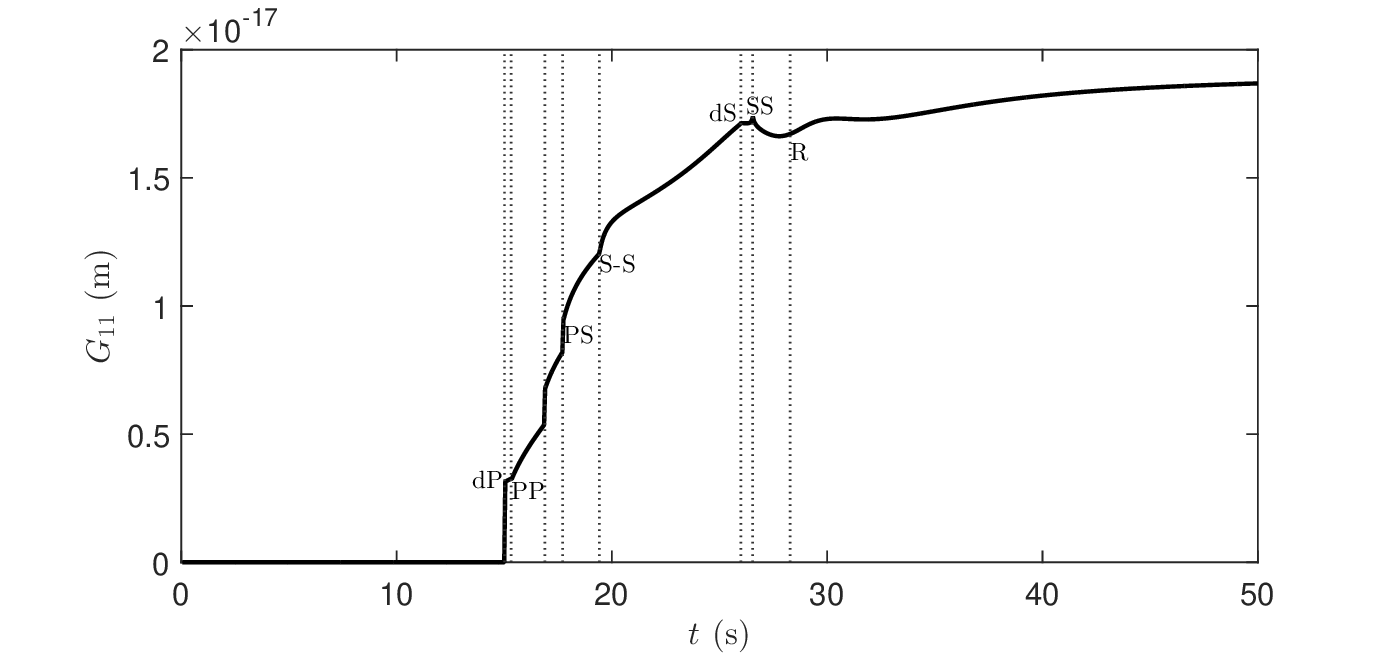}
\caption{
Time series from the full expressions of 
$G_{11}$, for a 1 Newton force with Heaviside step function time history at ($0$, $0$, $10$ km) observed at ($120$ km, $0$, $15$ km).
The eight vertical dotted lines mark the estimated arrival times of direct P, reflected PP, reflected SP, reflected PS, surface refracted S-S, direct S, reflected SS and Rayleigh waves, respectively.
The main motions of the wave field are associated with the arrival times of different types of the waves.
}
\label{Figure G11full}
\end{figure}

In Figure \ref{Figure G11full} time series from the full expressions of $G_{11}$ are shown. 
The main motions of the wave field correspond to the arrival times of different types of the waves. 
The SP wave arrives earlier that the PS wave because the source locates shallower than the receiver.
The disturbance of Rayleigh wave is obvious because the source and the receiver are quite shallow, compared to the distance between the source and the receiver. 

\begin{figure}
\centering
\subfigure{
\begin{minipage}{.48\textwidth}
\centering
\includegraphics[scale=0.6]
{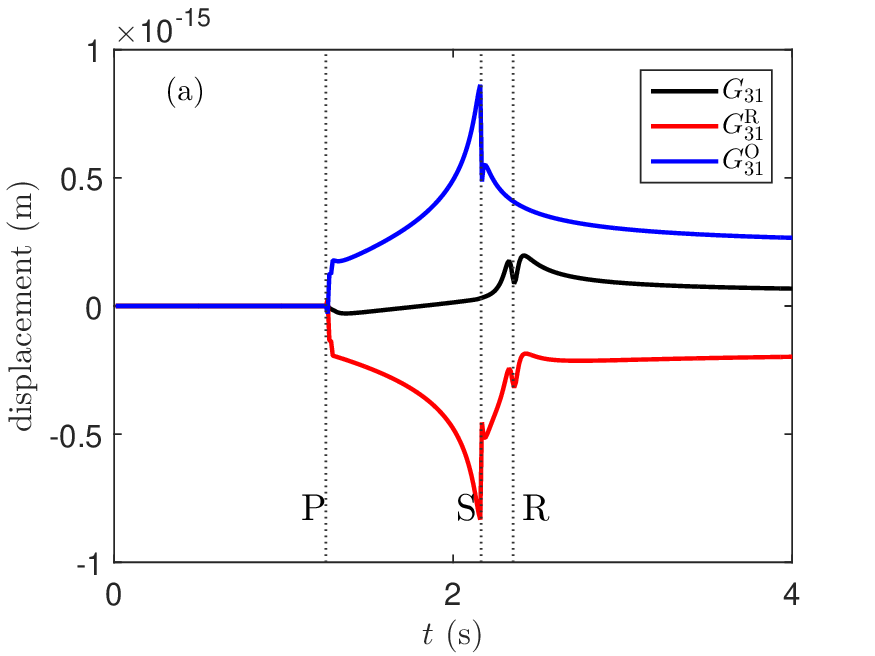}
\end{minipage}
}
\subfigure{
\begin{minipage}{.48\textwidth}
\centering
\includegraphics[scale=0.6]
{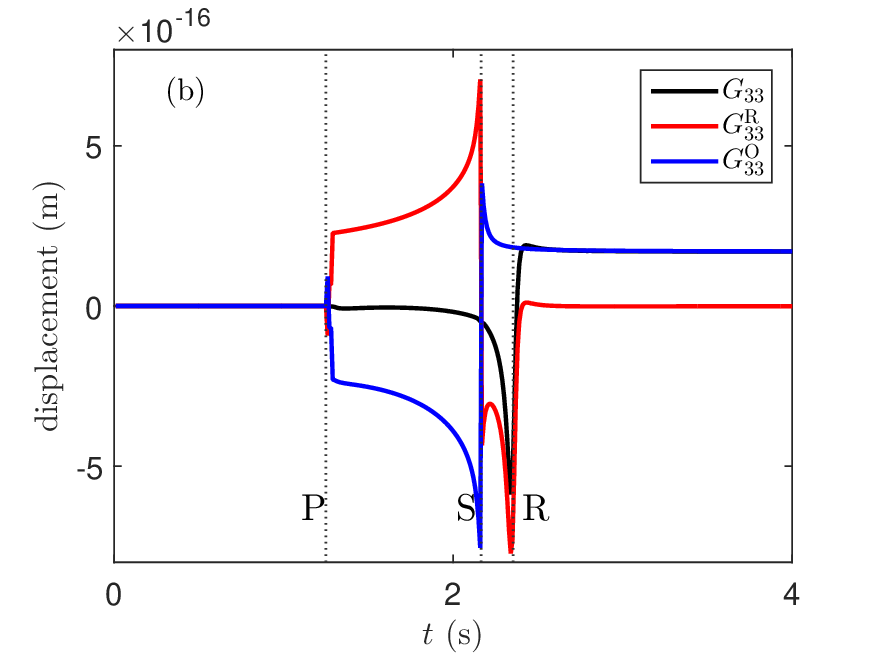}
\end{minipage}
}
\caption{
Time series from the full expressions of (a)
$G_{31}$ and (b) $G_{33}$ (black solid line), plotted together with the corresponding $G^{\rm R}_{ij}$ (red solid line) and $G^{\rm O}_{ij}$ (blue solid line). 
The force is set to 1 Newton with Heaviside step function time history at ($0$, $0$, $0.2$ km) observed at ($10$ km, $0$, $0.1$ km).
The three vertical dotted lines mark the estimated arrival times of direct P, direct S and Rayleigh waves.
The Rayleigh wave becomes the dominant phase after the arrival of S wave.
}
\label{Figure G31G33}
\end{figure}

To focus on the properties of the Rayleigh wave, we separate total tensor Green's function $\ts{G}$ into two parts:
\begin{align}
G_{ij}=G^{\rm R}_{ij}+G^{\rm O}_{ij},
\end{align}
where $G^{\rm R}_{ij}$ are known as the \textit{Rayleigh term} containing Rayleigh zeros, whereas $G^{\rm O}_{ij}$ are referred to as the ``Other'' term which is irrelevant to the Rayleigh wave. 
In Figure. \ref{Figure G31G33}, the time series of the full Green's function components $G_{31}$ and $G_{33}$ are shown, respectively, together with the corresponding time series of the Rayleigh and ``Other'' terms
$G^{\rm R}_{ij}$ and $G^{\rm O}_{ij}$.  

The numerical results indicate that the Rayleigh wave becomes the dominant phase when both the source and receiver are at shallow depths, compared to the range (i.e. to the epicentral distance). 
For the ``Other'' term, the displacements always decay smoothly after the arrival time of the direct S wave. 
For the \textit{Rayleigh term}, however, there is an obvious disturbance at the arrival time of the Rayleigh wave. 
Therefore, the \textit{Rayleigh term} can describe the complete behavior of the  Rayleigh wave.
We shall report on more details about the features of the Rayleigh wave in a forthcoming publication by the writers.
\section{Conclusion}\label{Section conclusion}
In this article, we derived the exact closed-form solution for the displacement in the interior of an
elastic half-space generated by a buried point force, which is referred to as Lamb's problem of the third kind. 
Our final expressions consist of only elementary functions and three kinds of standard elliptic functions, similar in form to those in FZ18, in which the response at a receiver located on the surface was considered. 
Our formulae reproduce exactly the same numerical results as those of J74, and our expressions have an obvious advantage over those of J74 when a particular kind of waves such as the Rayleigh wave is analyzed. 
On the basis of our results, the boundary integral equations (BIEs) can be established for the dynamic rupture propagation on a fault embedded in 3-D elastic half space. Zhang and Chen (2006) first extended the BIEs using the Green's function in half space, which is obtained by wavenumber integration method, and some numerical obstacles induced by the function are faced in their study. Our closed-form Green's function can be a great substitute to establish a more stable and efficient numerical algorithm.
Moreover, although the current formulae are valid when the Poisson's ratio is less than 0.2613, the derivation technique used here is still applicable, 
when the Poisson's ratio is beyond this value after minor modifications.
In a forthcoming companion publication, we will remove the restriction on the value of the Poisson's ratio and derive the closed-form solutions to Lamb's problem. 
More information about the Green's function in the half-space will be discussed based on our results.
\begin{acknowledgments}
This work was supported by the National Natural Science Foundation of China 
under grants 41874047 and 41674050 and by the High-performance Computing Platform of Peking University. 
We thank the two anonymous reviewers for their useful suggestions which are crucial to the improvement of our manuscript.
The authors acknowledge the course English Presentation for Geophysical Research of Peking University for help in improving this manuscript.
\end{acknowledgments}
\balance
 
\nobalance
\appendix
\section{Arrival time of the reflected PS wave}\label{Section tps}
\begin{figure}
\centering
\includegraphics[scale=0.9]
{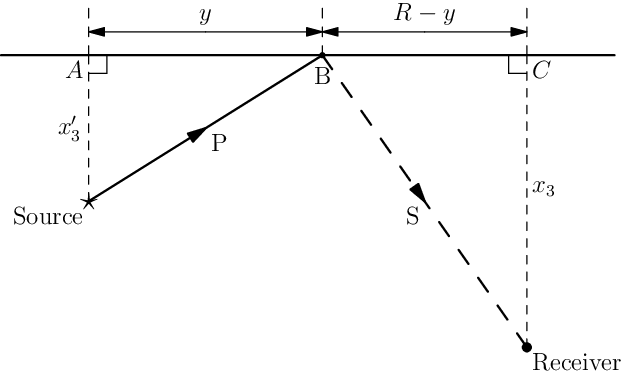}
\caption{
Schematic ray path for reflected PS wave. It goes at the speed of P wave from the location of the source to the point of incidence B on the free surface (solid line), then goes at the speed of S wave from B to the location of the receiver (dashed line). The source and the receiver are projected onto the surface as A and C, respectively. $y$ denotes the length of AB, $x'_3$, $x_3$ and $R$ have been shown in Figure \ref{Figure geometry}.
}
\label{tps}
\end{figure}

The ray path for reflected PS wave is shown in Figure \ref{tps}. Let $y=AB$, then the arrival time of the ray can be written as:
\begin{align}\label{A1}
t(y)=\sqrt{(x'_3)^2+y^2}/\alpha+\sqrt{x_3^2+(R-y)^2}/\beta.
\end{align} 
$t_{_{\rm PS}}$ is the minimum of the function $t(y)$, and the corresponding value $y_0$ is determined by:
\begin{align}\label{A2}
\left.\frac{\partial t(y)}{\partial y}\right|_{y=y_0}=0,
\end{align}
that is
\begin{align}\label{A3}
\frac{y_0}{\alpha\sqrt{(x'_3)^2+y_0^2}}=
\frac{R-y_0}{\beta\sqrt{(x_3)^2+(R-y_0)^2}}.
\end{align}
For convenience of calculation, eq. (\ref{A3}) can be written as a quartic equation:
\begin{small}
\begin{align}\label{A5}
\left(k^2-1\right)y_0^4-2R\left(k^2-1\right)y_0^3
+\left(R^2k^2+\left(x'_3\right)^2k^2-R^2-x^2_3\right)y_0^2
-2Rk^2\left(x'_3\right)^2y_0+R^2k^2\left(x'_3\right)^2=0,
\end{align}
\end{small}
where $0\leq y_0\leq R$. 
Therefore, $t_{_{\rm PS}}$ can be evaluated by eq. (\ref{A1}) together with the corresponding $y_0$ by eq. (\ref{A5}).
It is worth pointing out that $t_{_{\rm PS}}$ is the arrival time of the ray path satisfied the Snell's law. 
\section{Coefficients in the expressions for the reflected ${\rm PP}$ and ${\rm SS}$ waves}\label{Section coefficient}
The coefficients in the expressions for the reflected ${\rm PP}$ wave are determined through the partial fraction expansion as:
\begin{align}\label{GPP coeffient1}
\notag
\frac{A^{\rm PP}_1(B)M_{ij}^{\rm PP}(B)}{\sigma^{\rm PP}(B)}
=&\frac{u^{\rm PP}_{ij,1}}{B-\sigma^{\rm PP}_1}
+\frac{u^{\rm PP}_{ij,2}}{B-\sigma^{\rm PP}_2}
+\frac{u^{\rm PP}_{ij,3}}{B-\sigma^{\rm PP}_3}
+\frac{u^{\rm PP}_{ij,4}}{B-\sigma^{\rm PP}_4}
+\frac{u^{\rm PP}_{ij,5}+u^{\rm PP}_{ij,6}B}
{B^2+\left(\tilde{\sigma}_5^{\rm PP}\right)^2}
\\
&
+u^{\rm PP}_{ij,7}
+u^{\rm PP}_{ij,8}B
+u^{\rm PP}_{ij,9}B^2
+u^{\rm PP}_{ij,10}B^3
+u^{\rm PP}_{ij,11}B^4.
\end{align}
\begin{align}\label{GPP coeffient2}
\notag
\frac{A^{\rm PP}_2(B)M_{ij}^{\rm PP}(B)}{\sigma^{\rm PP}(B)}
=&\frac{v^{\rm PP}_{ij,1}}{B-\sigma^{\rm PP}_1}
+\frac{v^{\rm PP}_{ij,2}}{B-\sigma^{\rm PP}_2}
+\frac{v^{\rm PP}_{ij,3}}{B-\sigma^{\rm PP}_3}
+\frac{v^{\rm PP}_{ij,4}}{B-\sigma^{\rm PP}_4}
+\frac{v^{\rm PP}_{ij,5}+v^{\rm PP}_{ij,6}B}
{B^2+\left(\tilde{\sigma}_5^{\rm PP}\right)^2}
\\
&
+v^{\rm PP}_{ij,7}
+v^{\rm PP}_{ij,8}B
+v^{\rm PP}_{ij,9}B^2
+v^{\rm PP}_{ij,10}B^3
+v^{\rm PP}_{ij,11}B^4
+v^{\rm PP}_{ij,12}B^5.
\end{align}
where $\sigma_5^{\rm PP}={\rm i}\tilde{\sigma}_5^{\rm PP}$. 
$\sigma_i^{\rm PP}\ (i=1,\cdots,6)$ are the roots of the six-order polynomial:
\begin{align}
\sigma^{\rm PP}(B)=(2B^{2}+k^{2}-2)^{4}-16B^{2}(B^{2}+k^{2}-1)(B^{2}-1)^{2}.
\end{align}
When the Poisson's ratio is less than 0.2613, 
$\sigma_i^{\rm PP}\ (i=1,\cdots,4)$ are real, and 
$\sigma_5^{\rm PP}$ and $\sigma_6^{\rm PP}$ are purely imaginary.
$A^{\rm PP}_1(B)$, $A^{\rm PP}_2(B)$ and the 9 components of $M^{\rm PP}_{ij}(B)$ are:
\begin{align}\label{GPP co}
\notag
A^{\rm PP}_1(B)
=&(2B^2+k^2-2)^4+16B^2(B^2+k^2-1)(B^2-1)^2,
\\
\notag
A^{\rm PP}_2(B)
=&8B(B^2+k^2-1)(B^2-1)(2B^2+k^2-2)^2,
\\
\notag
M_{11}^{\rm PP}(B)=&\frac{\sin^2\phi-\cos^2\phi}{\sin^2\theta}(B\cos\theta-T_{\rm PP})^2
+\sin^2\phi(B^2-1),
\\
M_{22}^{\rm PP}(B)=&\frac{\cos^2\phi-\sin^2\phi}{\sin^2\theta}(B\cos\theta-T_{\rm PP})^2
+\cos^2\phi(B^2-1),
\\
\notag
M_{12}^{\rm PP}(B)=M_{21}^{\rm PP}(B)=&-\frac{2\cos\phi\sin\phi}{\sin^2\theta}(B\cos\theta-T_{\rm PP})^2-\cos\phi\sin\phi(B^2-1)
\\
\notag
M_{31}^{\rm PP}(B)=-M_{13}^{\rm PP}(B)=&\frac{\cos\phi}{\sin\theta}(B\cos\theta-T_{\rm PP})B,
\\
\notag
M_{32}^{\rm PP}(B)=-M_{23}^{\rm PP}(B)=&\frac{\sin\phi}{\sin\theta}(B\cos\theta-T_{\rm PP})B,
\\
\notag
M_{33}^{\rm PP}(B)=&B^2.
\end{align}
Similarly, for the reflected ${\rm SS}$ wave, we have
\begin{align}\label{GSS coeffient1}
\notag
\frac{M_{ij}^{\rm SS}(B)}{\sigma^{\rm SS}(B)}
=&\frac{u^{\rm SS}_{ij,1}}{B-\sigma^{\rm SS}_1}
+\frac{u^{\rm SS}_{ij,2}}{B-\sigma^{\rm SS}_2}
+\frac{u^{\rm SS}_{ij,3}}{B-\sigma^{\rm SS}_3}
+\frac{u^{\rm SS}_{ij,4}}{B-\sigma^{\rm SS}_4}
+\frac{u^{\rm SS}_{ij,5}+u^{\rm SS}_{ij,6}B}
{B^2+\left(\tilde{\sigma}_5^{\rm SS}\right)^2}
\\
&
+u^{\rm SS}_{ij,7}
+u^{\rm SS}_{ij,8}B
+u^{\rm SS}_{ij,9}B^2
+u^{\rm SS}_{ij,10}B^3
+u^{\rm SS}_{ij,11}B^4,
\end{align}
\begin{align}\label{GSS coeffient2}
\notag
\frac{N_{ij}^{\rm SS}(B)}{\sigma^{\rm SS}(B)}
=&\frac{v^{\rm SS}_{ij,1}}{B-\sigma^{\rm SS}_1}
+\frac{v^{\rm SS}_{ij,2}}{B-\sigma^{\rm SS}_2}
+\frac{v^{\rm SS}_{ij,3}}{B-\sigma^{\rm SS}_3}
+\frac{v^{\rm SS}_{ij,4}}{B-\sigma^{\rm SS}_4}
+\frac{v^{\rm SS}_{ij,5}+v^{\rm SS}_{ij,6}B}
{B^2+\left(\tilde{\sigma}_5^{\rm SS}\right)^2}
\\
&
+v^{\rm SS}_{ij,7}
+v^{\rm SS}_{ij,8}B
+v^{\rm SS}_{ij,9}B^2
+v^{\rm SS}_{ij,10}B^3
+v^{\rm SS}_{ij,11}B^4
+v^{\rm SS}_{ij,12}B^5.
\end{align}
where $\sigma_5^{\rm SS}={\rm i}\tilde{\sigma}_5^{\rm SS}$. 
$\sigma_i^{\rm SS}\ (i=1,\cdots,6)$ are the roots of the six-order polynomial:
\begin{align}
\sigma^{\rm SS}(B)=(2B^2-1)^4-16B^2(B^2+k^{-2}-1)(B^2-1)^2.
\end{align} 
The root distribution of $\sigma^{\rm SS}(B)$ is the same as those of 
$\sigma^{\rm PP}(B)$.
The 9 components of $M^{\rm SS}_{ij}(B)$ are:
\begin{align}\label{GSS co1}
\notag
M^{\rm SS}_{ij}(B)=&\left(2B^2-1\right)^2X_{ij}(B)
+16B^2\left(B^2-1\right)\left(B^2+k^{-2}-1\right)Y_{ij}(B),
\quad (i,j=1,2)
\\
M^{\rm SS}_{31}(B)=&-M^{\rm SS}_{13}(B)=\frac{\cos\phi}{\sin\theta}(B\cos\theta-T_{\rm SS})BA_1^{\rm SS}(B),
\\
\notag
M^{\rm SS}_{32}(B)=&-M^{\rm SS}_{23}(B)
=\frac{\sin\phi}{\sin\theta}(B\cos\theta-T_{\rm SS})BA_1^{\rm SS}(B),
\\
\notag
M^{\rm SS}_{33}(B)=&\left(B^2-1\right)A_1^{\rm SS}(B),
\end{align}
and the 9 components of $N^{\rm SS}_{ij}(B)$ are:
\begin{align}\label{GSS co2}
\notag
N^{\rm SS}_{ij}(B)=&4\left(B^2+k^{-2}-1\right)B\left(\left(B^2-1\right)X_{ij}(B)+\left(2B^2-1\right)^2Y_{ij}(B)\right),
\quad (i,j=1,2)
\\
N^{\rm SS}_{31}(B)=&-N^{\rm SS}_{13}(B)
=\frac{\cos\phi}{\sin\theta}(B\cos\theta-T_{\rm SS})BA_2^{\rm SS}(B),
\\
\notag
N^{\rm SS}_{32}(B)=&-N^{\rm SS}_{23}(B)
=\frac{\sin\phi}{\sin\theta}(B\cos\theta-T_{\rm SS})BA_2^{\rm SS}(B),
\\
\notag
N^{\rm SS}_{33}(B)=&\left(B^2-1\right)A_2^{\rm SS}(B),
\end{align}
where
\begin{align}\label{GSS co3}
\notag
A_1^{\rm SS}(B)=&(2B^2-1)^4+16B^2(B^2+k^{-2}-1)(B^2-1)^2,
\\
\notag
A_2^{\rm SS}(B)=&4(B^2+k^{-2}-1)B(B^2-1)(2B^2-1)^2,
\\
\notag
X_{11}(B)=&\sin^2\phi (B^2-1)(2B^2-1)^2+\frac{\sin^2\phi-\cos^2\phi}{\sin^2\theta}(B\cos\theta-T_{\rm SS})^2(2B^2-1)^2+(2B^2-1)^2,
\\
\notag
Y_{11}(B)=&\sin^2\phi(B^2-1)^2+
\frac{\sin^2\phi-\cos^2\phi}{\sin^2\theta}
(B^2-1)(B\cos\theta-T_{\rm SS})^2
-(2\cos^2\phi-1)(B^2-1)
\\
\notag
&
-2\frac{2\cos^2\phi-1}{\sin^2\theta}(B\cos\theta-T_{\rm SS})^2,
\\
X_{12}(B)=&-\frac{\cos\phi\sin\phi}{\sin^2\theta}\left(2B^2-1\right)^2
\left(
\left(B^2-1\right)\sin^2\theta
+2\left(B\cos\theta-T_{\rm SS}\right)^2
\right),
\\
\notag
Y_{12}(B)=&-\frac{\cos\phi\sin\phi}{\sin^2\theta}
\Big(
\left(B^2-1\right)^2\sin^2\theta
+2\left(B^2-1\right)(B\cos\theta-T_{\rm SS})^2
+2\left(B^2-1\right)\sin^2\theta
\\
\notag
&
+4\left(B\cos\theta-T_{\rm SS}\right)^2
\Big),
\\
\notag
X_{22}(B)=&\cos^2\phi (B^2-1)(2B^2-1)^2-\frac{\sin^2\phi-\cos^2\phi}{\sin^2\theta}(B\cos\theta-T_{\rm SS})^2(2B^2-1)^2+(2B^2-1)^2,
\\
\notag
Y_{22}(B)=&\cos^2\phi(B^2-1)^2-
\frac{\sin^2\phi-\cos^2\phi}{\sin^2\theta}
(B^2-1)(B\cos\theta-T_{\rm SS})^2
-(2\sin^2\phi-1)(B^2-1)
\\
\notag
&
-2\frac{2\sin^2\phi-1}{\sin^2\theta}(B\cos\theta-T_{\rm SS})^2.
\end{align}
\end{document}